\documentclass[longauth]{aa}  
\usepackage{graphicx}
\usepackage{epsfig,amssymb,amsmath,color}
\newcommand{\beq}{\begin{equation}}
\newcommand{\eeq}{\end{equation}}
\newcommand{\beqn}{\begin{eqnarray}}
\newcommand{\sinc}{\ {\rm sinc}}
\newcommand{\eeqn}{\end{eqnarray}}

\long\def\symbolfootnote[#1]#2{\begingroup%
\def\thefootnote{\fnsymbol{footnote}}\footnote[#1]{#2}\endgroup}
\usepackage{txfonts}
\usepackage{natbib}
\bibpunct{(}{)}{;}{a}{}{,}

\nonstopmode   

\begin{document}
   \title{Initial deep LOFAR observations of Epoch of Reionization windows: I. The North Celestial Pole}
\titlerunning{Initial deep LOFAR observations of EoR windows}

\author{S.~Yatawatta\inst{2}
  \and A.~G.~de~Bruyn\inst{1,2}
  \and M.~A.~Brentjens\inst{2}
  \and P.~Labropoulos\inst{2}
  \and V.~N.~Pandey\inst{2}
  \and S.~Kazemi\inst{1}
  \and S.~Zaroubi\inst{1}
  \and L.~V.~E.~Koopmans\inst{1}
  \and A.~R.~Offringa\inst{1,33}
  \and V.~Jeli\'c\inst{1}
  \and O.~Martinez Rubi\inst{1}
  \and V.~Veligatla\inst{1}
  \and S.~J.~Wijnholds\inst{2}
  \and W.~N.~Brouw\inst{2}
  \and G.~Bernardi\inst{12,1} % Harvard
  \and B.~Ciardi\inst{3}
  \and S.~Daiboo\inst{1}
  \and G.~Harker\inst{4}
  \and G.~Mellema\inst{5}
  \and J.~Schaye\inst{6}
  \and R.~Thomas\inst{1}
  \and H.~Vedantham\inst{1}
  \and E.~Chapman\inst{27}
  \and F.~B.~Abdalla\inst{27}
% LOFAR builders
  \and A.~Alexov\inst{28} % STSI
  \and J.~Anderson\inst{9} % Bonn
  \and I.~M.~Avruch\inst{10,1} % SRON + Kapteyn
  \and F.~Batejat\inst{29} % OSO
  \and M.~E.~Bell\inst{11,16} % Sydney
  \and M.~R.~Bell\inst{3} % Garching
  \and M.~Bentum\inst{2} % Astron
  \and P.~Best\inst{13} % Edinburgh
  \and A.~Bonafede\inst{14} % Bremen
  \and J.~Bregman\inst{2} % ASTRON
  \and F.~Breitling\inst{15} % Potsdam
  \and R.~H.~van de Brink\inst{2} % ASTRON
  \and J.~W.~Broderick\inst{16} % Southampton
  \and M.~Br\"uggen\inst{17,14} % Hamburg, Bremen
  \and J.~Conway\inst{29} % OSO
  \and F.~de Gasperin\inst{17} % Hamburg
  \and E.~de Geus\inst{2} % ASTRON
  \and S.~Duscha\inst{2} % ASTRON
  \and H.~Falcke\inst{20} % Radboud
  \and R.~A.~Fallows\inst{2} % ASTRON
  \and C.~Ferrari\inst{30} % Nice
  \and W.~Frieswijk\inst{2} % Astron
  \and M.~A.~Garrett\inst{2,6} % ASTRON+Leiden
  \and J.~M.~Griessmeier\inst{21,2} % CNRS
  \and A.~W.~Gunst\inst{2} % Astron
  \and T.~E.~Hassall\inst{16,22} % Southampton, Manchester
  \and J.~W.~T.~Hessels\inst{2,8} % Astron+UVA
  \and M.~Hoeft\inst{19} % Tautenburg
  \and M.~Iacobelli\inst{6} % Leiden
  \and E.~Juette\inst{18} % Bochum
  \and A.~Karastergiou\inst{23} % Oxford
  \and V.~I.~Kondratiev\inst{2,31}
  \and M.~Kramer\inst{9,22}
  \and M.~Kuniyoshi\inst{9} % Bonn
  \and G.~Kuper\inst{2}
  \and J.~van Leeuwen\inst{2,8}
  \and P.~Maat\inst{2}
  \and G.~Mann\inst{15} % Potsdam
  \and J.~P.~McKean\inst{2}
  \and M.~Mevius\inst{2,1}
  \and J.~D.~Mol\inst{2}
  \and H.~Munk\inst{2}
  \and R.~Nijboer\inst{2}
  \and J.~E.~Noordam\inst{2}
  \and M.~J.~Norden\inst{2} 
  \and E.~Orru\inst{2,20}
  \and H.~Paas\inst{32} % Groningen
  \and M.~Pandey-Pommier\inst{24,6} % Lyon + Leiden
  \and R.~Pizzo\inst{2}
  \and A.~G.~Polatidis\inst{2}
  \and W.~Reich\inst{9} % Bonn
  \and H.~J.~A.~R\"ottgering\inst{6} % Leiden
  \and J.~Sluman\inst{2}
  \and O.~Smirnov\inst{25} % RATT
  \and B.~Stappers\inst{22} % Manchester
  \and M.~Steinmetz\inst{15} % AIP
  \and M.~Tagger\inst{21} % CNRS
  \and Y.~Tang\inst{2} % Astron
  \and C.~Tasse\inst{7} % Paris
  \and S.~ter Veen\inst{20}
  \and R.~Vermeulen\inst{2}
  \and R.~J.~van~Weeren\inst{6,2,12}
  \and M.~Wise\inst{2}
  \and O.~Wucknitz\inst{26,9} % Bonn2 + Bonn3 + Bonn
  \and P.~Zarka\inst{7}
}

\institute{
% 1
  University of Groningen, Kapteyn Astronomical Institute, PO Box 800, 9700 AV Groningen, The Netherlands. 
%\\\email{ger@astron.nl}
% 2
  \and ASTRON, PO Box 2, 7990 AA Dwingeloo, The Netherlands. \email{yatawatta@astron.nl}
% 3
  \and Max-Planck Institute for Astrophysics, Karl-Schwarzschild-Strasse 1, 85748 Garching bei M\"unchen, Germany.
% 4
  \and Center for Astrophysics and Space Astronomy, University of Colorado, 389 UCB, Boulder, Colorado 80309-0389, USA.
% 5
  \and Department of Astronomy \& Oskar Klein Centre, AlbaNova, Stockholm University, SE-10691 Stockholm, Sweden
% 6
  \and Leiden Observatory, Leiden University, PO Box 9513, 2300 RA Leiden, The Netherlands.
% 7
  \and Observatoire de Paris, FR 92195 Meudon, France.
% 8
  \and University of Amsterdam, Astronomical Institute Anton Pannekoek, PO Box 94249, 1090 GE Amsterdam, The Netherlands.
% 9
  \and Max-Planck Institute for Astrophysics, P.O. Box 20 24, D-53010 Bonn, Germany.
% 10
  \and SRON Netherlands Institute for Space Research, PO Box 800, 9700 AV Groningen, The Netherlands.
% 11
  \and Sydney Institute for Astronomy, School of Physics A28, University of Sydney, NSW 2006, Australia.
% 12
  \and Harvard-Smithsonian Center for Astrophysics, 60 Garden Street, Cambridge, MA 02138, USA.
% 13
  \and Royal Observatory Edinburgh, Blackford Hill, Edinburgh, EH9 3HJ, UK.
% 14
  \and Jacobs University Bremen, Campus Ring 1, 28759 Bremen, Germany.
% 15
  \and Astrophysical Institute Potsdam, An der Sternwarte 16, 14482 Potsdam, Germany.
% 16
  \and University of Southampton, University Road, Southampton SO17 1BJ, UK.
% 17
  \and University of Hamburg, Gojenbergsweg 112, 21029 Hamburg, Germany.
% 18
%  \and Chalmers University of Technology, SE-412 96 Gothenburg, Sweden.
% 18
  \and Astronomisches Institut der Ruhr-Universit\"{a}t Bochum, Universitaetsstrasse 150, 44780 Bochum, Germany.
% 19
  \and Th\"uringer Landessternwarte, Tautenburg Observatory, Sternwarte 5, D-07778 Tautenburg, Germany.
% 20
  \and Radboud University Nijmegen, Faculty of NWI, PO Box 9010, 6500 GL Nijmegen, The Netherlands.
% 21
  \and Centre national de la recherche scientifique, 3 rue Michel-Ange, 75794 Paris cedex 16, France.
% 22
  \and Jodrell Bank Center for Astrophysics, School of Physics and Astronomy, The University of Manchester, Manchester M13 9PL,UK.
% 23
  \and University of Oxford, Wellington Square, Oxford OX1 2JD, UK.
% 24
  \and Centre de Recherche Astrophysique de Lyon, Observatoire de Lyon, 9 av
Charles Andr\'e, 69561 Saint Genis Laval Cedex, France.
% 25
  \and Rhodes University, RATT, Dep. Physics and Electronics, PO Box 94, Grahamstown 6140, South Africa.
% 27
%  \and University of Bonn, Regina-Pacis-Weg 3, D-53012 Bonn, Germany.
% 26
  \and Argelander-Institut f\"ur Astronomie, Auf dem H\"ugel 71, 53121 Bonn, Germany.
% 27
  \and UCL, Department of Physics \& Astronomy, University College London, Gower
Street, London, WC1E 6BT, UK.
% 28
  \and Space Telescope Science Institute, 3700 San Martin Drive, Baltimore, MD 21218, USA.
% 29
\and Onsala Space Observatory, Dept. of Earth and Space Sciences, Chalmers University of Technology, SE-43992 Onsala, Sweden.
% 30
\and Laboratoire Lagrange, UMR7293, Universit\`{e} de Nice Sophia-Antipolis, CNRS, Observatoire de la C\'{o}te d'Azur, 06300 Nice, France. 
% 31
\and Astro Space Center of the Lebedev Physical Institute, Profsoyuznaya str. 84/32, Moscow 117997, Russia. 
% 32
\and Center for Information Technology (CIT), University of Groningen, The Netherlands.
% 33
  \and Mount Stromlo Observatory, RSAA, Cotter Road, Weston Creek, ACT 2611, Australia. 
}

   \date{Received}

 \abstract
% context
{}
% aims
{
The aim of the LOFAR Epoch of Reionization (EoR) project is to detect the  spectral fluctuations of the redshifted HI 21cm signal. This signal is weaker by several orders of magnitude than the astrophysical foreground signals and hence, in order to achieve this, very long integrations, accurate calibration for stations and ionosphere and reliable foreground removal are essential. 
}
% methods
{
One of the prospective observing windows for the LOFAR EoR project will be centered at the North Celestial Pole (NCP).
We present results from  observations of the NCP window using the LOFAR highband antenna (HBA) array in the frequency range 115 MHz to 163 MHz. The data were obtained in April 2011 during the commissioning phase of LOFAR. We used baselines up to about 30 km.  The data was processed using a dedicated processing pipeline which is an enhanced version of the standard LOFAR processing pipeline.  
}
% results
{
With about 3 nights, of 6 hours each, effective integration we have achieved a noise level of about 100 $\mu$Jy/PSF in the NCP window. Close to the NCP, the noise level increases to about 180 $\mu$Jy/PSF, mainly due to additional contamination from unsubtracted nearby sources. We estimate that in our best night, we have reached a noise level only a factor of 1.4 above the thermal limit set by the noise from our Galaxy and the receivers. Our continuum images are several times  deeper than have been achieved previously using the WSRT and GMRT arrays.  We derive an analytical explanation for the excess noise that we believe to be mainly due to sources at large angular separation from the NCP. We present some details of the data processing challenges and how we solved them. 
}
% conclusions
{
Although many LOFAR stations were, at the time of the observations, in a still poorly calibrated state we have seen no artefacts in our images which would prevent us from producing deeper  images in much longer  integrations on the NCP window which are about to commence. The limitations present in our current results are mainly due to sidelobe noise from the large number of distant sources, as well as errors related to station beam variations and rapid ionospheric phase fluctuations acting on bright sources. We are confident that we can improve our results with refined processing.
} 
% 5 {} token are mandatory

\maketitle
%
%________________________________________________________________
\section{Introduction}
A major epoch in the history of the Universe yet to be understood in detail is its Dark Ages and the Epoch of Reionization (EoR). Observational evidence for this era can be gathered with high probability by studying the fluctuations of the redshifted neutral hydrogen at redshifts corresponding to $6<z<12$. Therefore, there are numerous experiments becoming operational and already collecting data, especially in the frequency range from 115 MHz to 240 MHz to reach this goal.

At the forefront of such experiments is the Low Frequency Array (LOFAR) \citep{LOFARp}. Similar EoR experiments using other radio telescopes are already underway. For instance, \cite{GMRTUpper} provide a new lower bound for the statistical detection threshold of HI fluctuations using the GMRT (Giant Metrewave Radio Telescope). While MWA (Murchison Widefield Array) and PAPER (Precision Array to Probe the EoR) are not in full hardware deployment yet, there are still relevant results being produced. In \cite{MWA32T} and \cite{MWAWilliams}, initial widefield images of the southern sky using 32 MWA Tiles are presented. In \cite{PAPERSource}, full sky images and source catalogs using PAPER are presented. 

In preparing for the LOFAR EoR project we have conducted several pilot
experiments with the Low Frequency Frontends on the WSRT in a relevant
frequency range: 138-157 MHz. The results of these observations, and a
discussion of their limitations, have been described by \citet{Ber1,Ber}.  For LOFAR in its commissioning phase we have adopted a
multi-faceted observing strategy, building on the experience gained
from the WSRT data. The rationale behind this is described in more detail 
in \citet{EORp}. A brief summary follows:

Using LOFAR in its commissioning phase we have observed and processed three 
very diverse windows. One window contains a very bright compact source, 
3C196, which allows exquisite absolute calibration as well as a study of the 
systematics at very high spectral and image dynamic range. At a 
declination of only 48 degrees it will allow a study of  elevation 
dependent effects. The many bright compact field sources around 3C196 also 
allow a study of  ionospheric non-isoplanaticity.  The results of these
observations, with emphasis on all those topics, will be described in a
separate paper \citep{3C196p}.  The second window was chosen to ascertain possible damaging effects of faint signals due to instrumental leakage of bright polarized Galactic foreground signals. These results, on the Elais N1 window,  will be described by \citet{Eliasp} in the second paper in this series.

The third window, without bright sources and with relatively faint diffuse linearly polarized
emission from our Galaxy, is accessible throughout the year. This window is centered
on the North Celestial Pole (NCP). It has been used to experiment with
various calibration approaches as well as conduct a thorough analysis of the
noise levels attainable with the current, still incomplete, LOFAR
array.  These results are the subject of the first paper in the
series. 

The three windows described above have thus far been observed using a single digital
beam with about 48 MHz bandwidth.  Between these three windows we expect
to address most of the issues that will affect much longer observations with
LOFAR, which should go one order of magnitude deeper in noise level.  The 
analysis of  the results obtained in these three windows, as they pertain 
to EoR  signal levels, will be discussed in more detail in subsequent 
publications.

In this paper, we present results of LOFAR observations pointed at the NCP in the frequency range 115 MHz to 163 MHz. The NCP was previously observed using the Westerbork Synthesis Radio Telescope (WSRT) in a similar frequency range, albeit with limited integration time and resolution. As reported by \cite{Ber}, the WSRT observations are mainly limited by broadband (and low level) radio frequency interference (RFI), and classical confusion (due to having limited $<$ 3km longest baseline) that prevents reaching the theoretical noise level.

LOFAR provides significant challenges as well as advantages over conventional low frequency radio telescopes in terms of calibration. Unlike the WSRT, antennas on the ground are much less susceptible to broadband RFI. On the other hand, calibration of a LOFAR observation is challenging due to many reasons including spatially and temporally varying beam shapes with wide fields of view as well as  mild to severe ionospheric distortions.  Therefore, it is paramount that we test and demonstrate the feasibility of LOFAR for EoR observations before starting the long, dedicated observing campaign.

The results reported in this paper are based on integrations of the NCP which consisted of 3 nights with 6 hours each nighttime observing. We provide details of the calibration and imaging that lead us to almost reach the expected theoretical noise level (within a factor of 1.4). We also provide details of current limitations and what we expect with the current, still incomplete LOFAR. Based on this result, we see no show stoppers for the launch of the dedicated LOFAR EoR observations on this window which will last several hundred hours.

This paper is organized as follows: In section \ref{sec:obs}, we give an extensive overview of the observational setup. Section \ref{sec:calib} describes the data processing pipeline. In section \ref{sec:results}, we present initial results with deep very widefield images, new sources, and the noise behaviour. We give an analytical explanation for the noise behavior in section \ref{sec:outlier} that considers the excess noise due to sources far away from the phase center. Finally, we draw our conclusions in section \ref{sec:conc}.

Notation (mostly in section \ref{sec:outlier}): We use bold lowercase letters for vectors and bold uppercase letters for matrices. The matrix Frobenius norm is given by $\|.\|$. The matrix transpose, Hermitian transpose and pseudoinverse are given by $(.)^T$, $(.)^H$ and $(.)^{\dagger}$, respectively. The trace of a matrix is given by $\mathrm{trace}(.)$. The real part of a complex number is denoted by $\mathrm{Re}(.)$. The statistical expectation operator is denoted by $E\{.\}$.
\section{Observational Setup\label{sec:obs}}
In this section we provide details of the LOFAR stations used in the NCP observations. We also provide the motivation behind observing the NCP.
\subsection{LOFAR stations}
We give a brief overview of LOFAR hardware and a complete overview can be found in \citet{LOFARp} and \citet{EORp}. Each LOFAR HBA station consists of multiple elements (dipoles) with dual, linear polarized receivers. For a core station (CS), there are 384 dipoles, arranged in 24 tiles that have dipoles on a $4\times4$ grid. For a remote station (RS), there are 768 dipoles, arranged in 48 tiles. The signals of each dipole in a tile are coherently added (or beamformed) to form a narrow field of view (FOV) along a given direction in the sky. The effective beam shape is the product of the array (beamformer) gain with the dipole beam shape. It should be noted that the dipole beam shape is strongly polarized, and along some directions in the sky the polarization could be as much as 20\% of the total intensity.

The nominal LOFAR FOV at around 150 MHz at the NCP (from null to null) is about 11 degrees in diameter for a CS and about 8 degrees in diameter for a RS. Therefore, the effective FOV is about 10 degrees in diameter. There is also a complicated  low level sidelobe pattern surrounding the FOV, and the sidelobes change with time and frequency, as the beamformer weights change. In order to minimize the cumulative effect of the sidelobes, each LOFAR station is given a different rotation in its dipole layout (but keeping the dipoles parallel). Due to this reason, each LOFAR station will have a unique beam shape, that varies in time, frequency as well as according to the direction being pointed at. For a widefield image, this naturally leads to beam variations that depend on time, frequency as well as the direction in the sky.

\subsection{The NCP and its surroundings}
\begin{figure}[htbp]
\begin{center}
\epsfxsize=3.4in \leavevmode\epsfbox{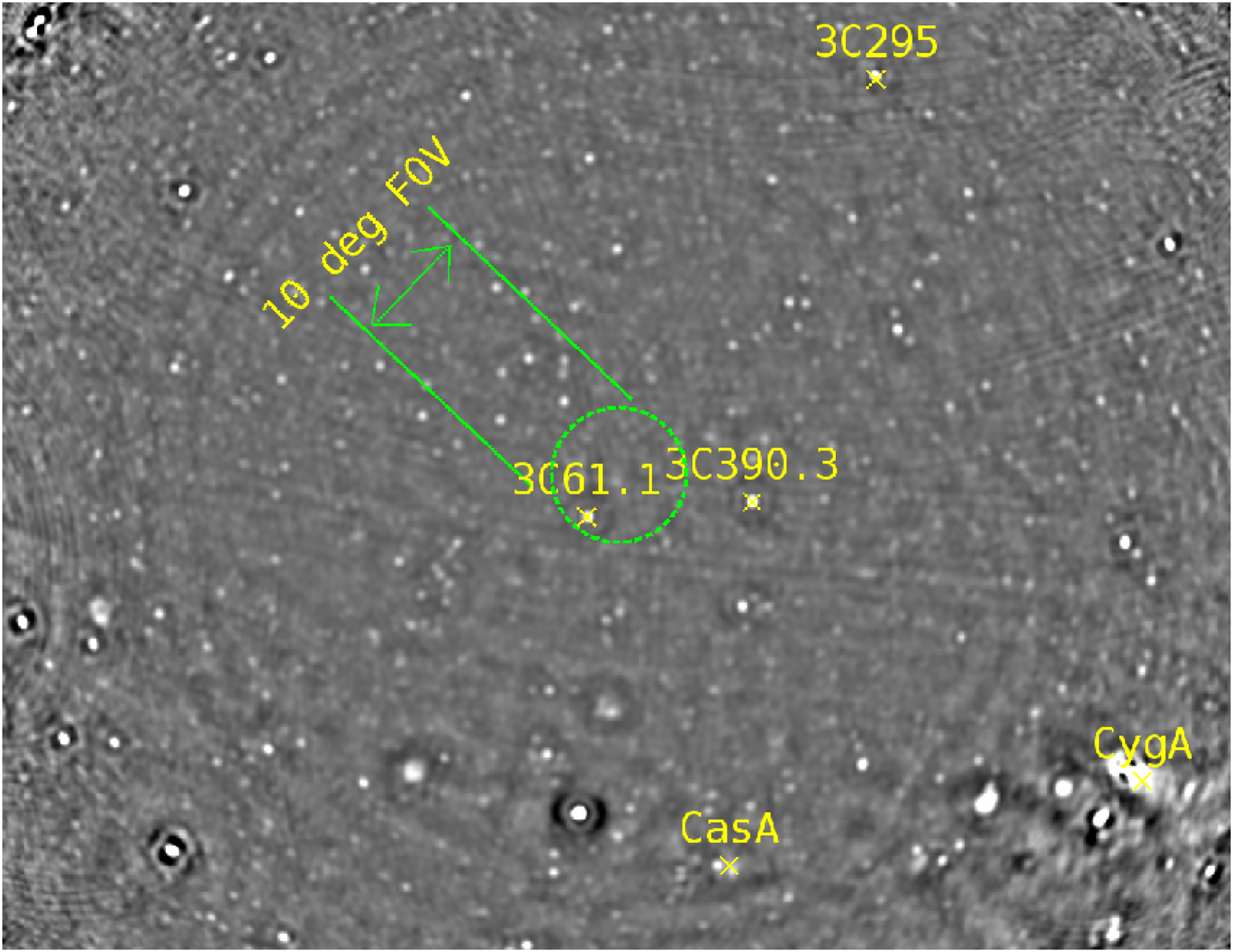}
\caption{NCP FOV overlaid on a full sky image at 50 MHz. The full sky image is made using the LOFAR lowband dipoles.\label{FOV}}
\end{center}
\end{figure}

It is of significant importance that the NCP FOV lies on a relatively cold (i.e., having low sky temperature) spot in the Galactic halo in order to minimize the effects of Galactic foregrounds. In Fig. \ref{FOV}, we show the NCP window (or FOV), which is overlaid on a full sky image observed at 50 MHz. The full sky image was made using an array of 16 LOFAR lowband dipoles, with the longest baseline of 450 m.

\begin{figure}[htbp]
\begin{center}
\epsfxsize=2.4in \leavevmode\epsfbox{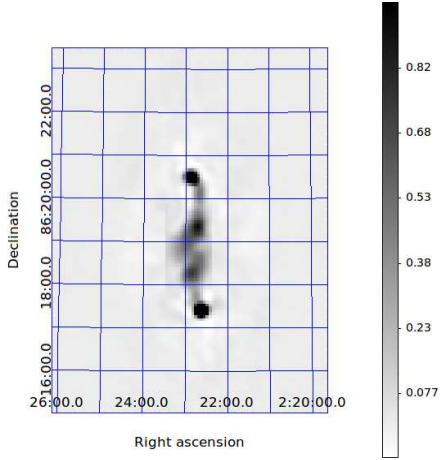}
\caption{3C61.1 model image at 150 MHz. The two hot spots in this source, a giant double radio galaxy at z=0.188 \citep{3C61red}, have peak flux densities of about 7 Jy in an 8$^{\prime\prime}$ PSF. The colourbar units are in Jy/PSF.\label{3c61}}
\end{center}
\end{figure}

The Galactic plane lies along an arc joining Cassiopeia A (CasA) and Cygnus A (CygA) in Fig. \ref{FOV} but it is resolved. CasA is 32 degrees away from the pole while CygA is about 40 degrees away. The closest 3C source to the pole is 3C61.1 which is 4 degrees away, with a total flux about 35 Jy (peak about 7 Jy) at 150 MHz. It is fully resolved and an image made from a previous LOFAR observation is shown in Fig. \ref{3c61}. There are several other bright 3C sources in the vicinity including 3C390.3 (11 degrees away).

\begin{figure}[htbp]
\begin{center}
\epsfxsize=3.0in \leavevmode\epsfbox{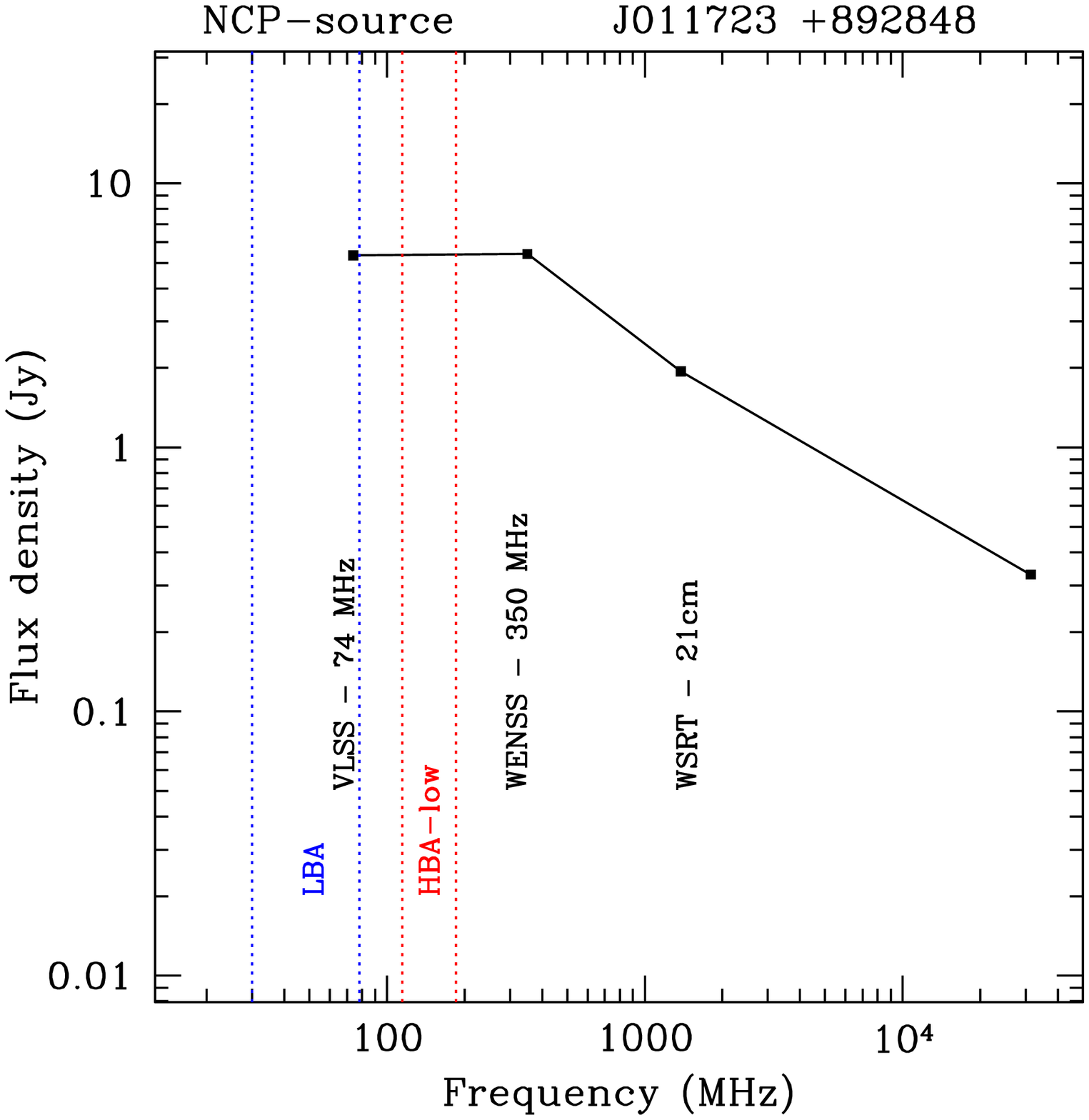}
\caption{Radio spectrum of  NVSS J011732+892848 compiled from data in the literature \citep{WENSS,VLSS}.}
\label{SpecI}
\end{center}
\end{figure}

Due to the station beams, many of the strong sources outside the main beam are suppressed significantly and the brightest (apparent) source in the FOV is NVSS J011732+892848,  about 30$^\prime$ away from the pole. This source is unresolved in our observations and has a peak flux of 5.4 Jy at 352 MHz \citep{WENSS} and a peak flux of 5.3 Jy at 74 MHz  \citep{VLSS} as reported in the WENSS and VLSS surveys, respectively. Therefore, we assume this source to have a flat spectrum within the observing band as in Fig. \ref{SpecI}, which makes it a suitable candidate for absolute flux calibration and noise estimation. Because it is very close to the pole, the nominal station beam gain is equal to the beam gain along the direction of the NCP (which is unity) and constant in time. However, the element (dipole) gain varies and this is taken into account during calibration.

\subsection{Motivation for observing the NCP}
The NCP is one of several observational windows for LOFAR EoR observations \citep{EORp}. The reasons  behind choosing the NCP are numerous although this does not imply it to be the optimal choice. We list some of the positive and some of the potentially negative aspects of this choice: 
\begin{itemize}
\item[+] The geographical location of LOFAR makes the NCP window observable at night time, throughout the year, at high elevation (53 degrees).
\item[+] Due to minimum projection effects of the $uv$ tracks, we get almost circular $uv$ coverage and therefore, we get an almost circular point spread function (PSF).
\item[$\pm$] The strongest source in the FOV, 3C61.1 is attenuated by about 70\% and the strongest (apparent) source is about 5 Jy in peak flux. Not having a strong source in the FOV has both advantages and disadvantages. First, not having a strong source means a not so high signal to noise ratio (SNR) in calibration. However, there are less artefacts resulting from deconvolution residuals of strong sources.
\item[$\pm$] Geostationary RFI that manages to escape flagging routines, which work on high-noise samples, may end up near the North Celestial Pole. Nonetheless, this provides a sensitive diagnostic of the presence of any faint, stationary, undetected  RFI.
\item[-] Finally, the NCP is located at a Galactic latitude of only 38 degrees. The overall system noise is therefore higher than the coldest regions near the North Galactic Pole.
\end{itemize}

\subsection{Observational parameters}
\begin{figure}[htbp]
\begin{center}
\epsfxsize=3.4in \leavevmode\epsfbox{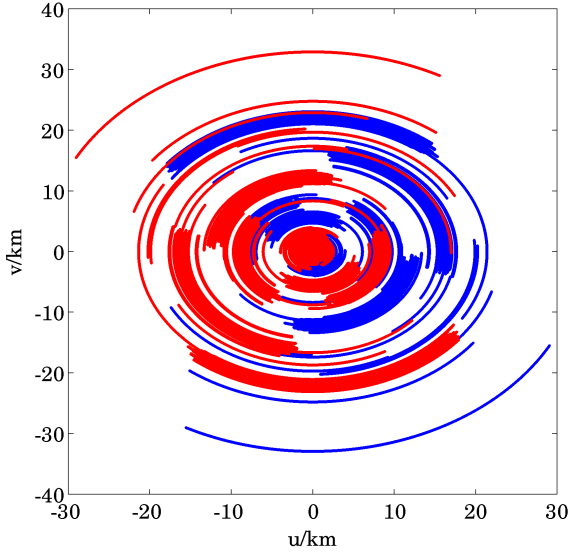}
\caption{6 hour monochromatic uv coverage at the NCP, using 40 core stations and 7 remote stations. The longest baseline is about 30 km. The two (red and blue) colours show the symmetric uv points obtained by conjugation of the data.\label{uvcov}}
\end{center}
\end{figure}

We used 40 core stations and 7 remote stations in our observations. The shortest baseline is 60 m and the longest baseline is about 30 km. The observing frequency range is from 115 MHz to 163 MHz. There are about 240 subbands in each observation, within this observing frequency range. Each subband has 256 channels, covering a bandwidth of 195 kHz. The monochromatic $uv$ coverage for a typical 6 hour NCP observation is shown in Fig. \ref{uvcov}.

With this $uv$ coverage, we get a resolution of about 12$^{\prime\prime}$ at 150 MHz. 
For comparison, in previous NCP observations using the WSRT \citep{Ber}, the longest baseline used was only 2.7\,km yielding an angular  resolution of only 120$^{\prime\prime}$. 
The  correlator integration time is set at 2 s.  We used data taken on 3 different nights for the results presented in this paper. In Table \ref{tab:results}, we summarize the observational parameters for these 3 different nights.

\begin{table*}[htbp]
\centering
\begin{tabular}{|c|c|c|c|c|}
\hline
Observation No. & Start Time (UTC) & No. Stations & No. Subbands & Noise ($\mu$Jy PSF$^{-1}$)\\ \newline 
& (duration 6 hours) & (delivering good data) & (processed) & \\
\hline
L24560 & 27-March-2011 20:00:05 & 45 & 229 & 125\\
L25085 & 10-April-2011 20:00:05 & 43 & 185 & 255\\
L26773 & 19-May-2011 19:30:00 & 41 & 187 & 224\\
\hline
\end{tabular}
\caption{Summary of observational parameters (and noise level achieved with uniformly weighted images, at the edge of the FOV) for the 3 nights of data taken. The variability of the noise is due to variability in the sensitivity of the stations, some of which were less focused due to beamforming errors at the time of the observations.}
\label{tab:results}
\end{table*}

\section{Data Reduction\label{sec:calib}}
In this section, we describe the major steps taken to calibrate the NCP observations. Apart from the initial processing, the data was completely processed in a CPU/GPU\footnote{CPU: Central Processing Unit, GPU: Graphics Processing Unit} cluster dedicated for LOFAR EoR computing needs. The processing of these observations also enabled us to fine tune the software used in various processing steps.

\subsection{Initial processing}
The LOFAR correlator \citep{Romein:10} outputs data at a very fine resolution (2 s and 0.78 kHz), mainly to facilitate RFI mitigation. However, the data volume makes it cumbersome for further processing. Therefore, the  data of each subband (having 256 channels at 2 s integration) is flagged using the  {\tt aoflagger} \citep{aoflagger,AOflagger2} and averaged to 15 channels (after removing the 8 channels at both subband  edges, mainly to remove edge effects from the polyphase filter). This significantly reduces the size of the data that has to be processed in the following stages. However, with improvements in software, we intend to process data at a finer frequency resolution, once regular EoR observing has begun.

\subsection{Sky model}
Because there is no single bright source in the FOV, the sky model used for initial calibration of the NCP data contains about three hundred discrete sources, spread across the FOV of about 10$\times$10 square degrees.
The most complex source in this region is 3C61.1 as shown in Fig. \ref{3c61}. In order to efficiently model this source, we use a model including shapelets \citep{SYURSI2011} and point sources. The rest of the sky model is modeled as a set of discrete sources, having multiple point source components. The brightest source  (NVSS J011732+892848) is modeled as a single point source with a flat spectrum and a peak flux of 5.3 Jy. We like to emphasize here that we diverge from the traditional 'clean component' based sky model construction in order to minimize the number of components used without the loss of accuracy \citep{SAM2010}. In order to automate this process, we have developed custom software ({\tt buildsky}) that creates a sky model with the minimum number of source components required. The principle behind {\tt buildsky} is to select the simplest model for a given source by choosing the correct number of degrees of freedom \citep{SYURSI2011}. While a point source has only one degree of freedom (for its shape), a double source has two and so on. There are additional degrees of freedom due to its position and flux. We use information theoretic criteria as given in \citet{SYURSI2011} to select the optimum number of degrees of freedom for any given source.
All the sources in the sky model are unpolarized. The sky model was updated using two calibration and imaging cycles.
\subsection{Calibration}
The aim of calibration of LOFAR EoR observations is twofold: (i) correction for instrumental and ionospheric errors in the data (ii) removal of strong foreground sources from the data such that specialized foreground removal algorithms (e.g. \cite{Harker}) can be applied. The basic description of the LOFAR EoR data model used in calibration is given by \cite{Panos}. We use an enhanced version of the LOFAR calibration pipeline \citep{BBS} for the EoR data calibration.

\subsubsection{Data correction\label{calib_corr}}
Major steps in our calibration pipeline are as follows:
\begin{enumerate}
\item We first calibrate for clock errors as well as small time scale ionospheric errors along the center of the FOV. This is the so called $uv$ plane or direction independent calibration \citep{Panos} and is performed using the Black Board Selfcal ({\tt BBS}) package \citep{BBS}. At this stage, each subband has 15 channels at 2~s integration time. We determine the calibration solutions for every 10~s, and one solution per subband. Since we do not have a dominant source at the center of the FOV, the solutions thus obtained correspond to small time scale ionospheric phase fluctuations common to the full FOV plus the clock errors.
\item Once $uv$ plane calibration is done, the data is corrected for these errors. The data is also  corrected for the element beam gain along the center of the FOV. As discussed previously, the dipole beam of LOFAR is strongly polarized and we use an element beam model based on numerical simulations. For an area about 10 degrees in diameter in the sky, the variation of the dipole beam shape is assumed to be small and correction for the center of the FOV is considered accurate enough for the full FOV.
\item The corrected data is averaged to 183 kHz (one channel per subband) and 10 s integration time. The data is also flagged by clipping any spikes present in the data after correction.
\end{enumerate}

\subsubsection{Source subtraction\label{calib_sub}}
LOFAR has a very wide field of view and along each direction, the errors present in the data are different due to varying beam shape and ionospheric effects \citep{Koopmans}. Therefore, source subtraction is not a simple deconvolution problem for LOFAR observations. Even for a simple deconvolution, it is better to subtract the sources directly from the visibilities \citep{SAM2010}. In the case of LOFAR, this subtraction has to be done with the appropriate gain corrections along each direction.

\begin{figure}[htbp]
\begin{center}
\epsfxsize=3.2in \leavevmode\epsfbox{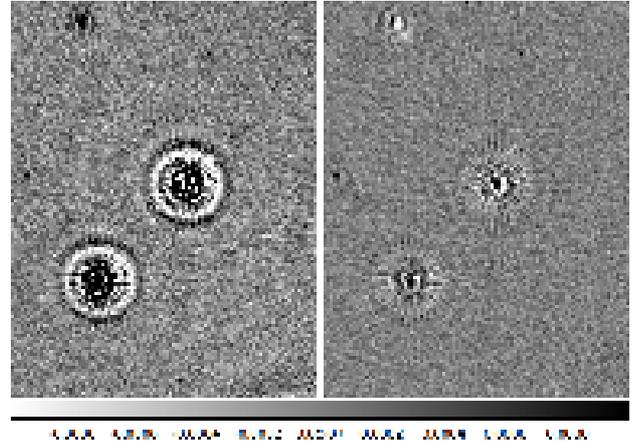}
\caption{Clustering of two sources 5$^\prime$  apart. The left panel shows the two (point) sources with identical error patterns. The right panel shows the image made after determining a common error (at an interval of 20 minutes) for both sources and subtracting their contribution from the data. The sky model has been restored in the right panel. The colourbar (bottom) units are in Jy/PSF.\label{ncp_cluster_two}}
\end{center}
\end{figure}

In order to efficiently and accurately solve the multi-source calibration problem, we have developed algorithms and software based on Expectation Maximization \citep{SAGE,Kaz2,SAGECAL}. We have implemented these algorithms ({\tt SAGECal}) with accelerated processing using graphics processing units (GPUs). In the NCP window, there are about 500 bright discrete sources (note that we subtract more sources than what we use for the $uv$ plane calibration, for which we only use about 300) that are subtracted from calibrated visibilities with the correct directional gains. We have 'clustered' \citep{Kaz1} these sources into about 150 different directions. Thus we  effectively determine the errors along 150 directions during the source subtraction.  An example of clustering is shown in Fig. \ref{ncp_cluster_two}. In the left panel of Fig. \ref{ncp_cluster_two}, we show two sources that are about 5$^\prime$ apart and apparently having identical error patterns. Therefore, instead of calibrating along each source individually, we can cluster them into one complex source and determine the common errors. The right panel in Fig. \ref{ncp_cluster_two} shows the result after subtracting the cluster and restoring the model.

There are still some errors remaining in the right panel of Fig. \ref{ncp_cluster_two} mainly because of errors in the sky model and due to the effect of surrounding sources (that were not included in the sky model) and also due to short time scale ionospheric errors. It should also be mentioned that while most of the sources subtracted lie within the FOV, we have also subtracted strong sources far away from the NCP as shown in Fig. \ref{FOV}, for example CasA.

\subsection{Imaging}
We make images at different stages during calibration. All images are made using CASA \footnote{Common Astronomy Software Applications, http://casa.nrao.edu}. In order to update the sky model, we make images of the calibrated and source subtracted data. We keep the highest available resolution in order to create accurate source models. For the results presented in this paper, we have a resolution of about 12$^{\prime\prime}$ and we choose a pixel size of 4$^{\prime\prime}$ with uniform weighting. Even though the nominal FOV is about 10 degrees, we make images that have an FOV of about 13 degrees, to detect sources at the edge of the beam. We restore the subtracted sources onto these images, after convolving with the nominal (Gaussian) PSF. Afterwards, we use {\tt Duchamp} \citep{Duchamp} to select areas with positive flux and update the sky model using {\tt buildsky}.

More relevant for EoR signal detection are images made at low resolution using the short baselines of LOFAR. Therefore, we also make images using baselines less than 1200 wavelengths at 35$^{\prime\prime}$ pixel size. The image size is chosen to be about 65 degrees so that we can see any contributions from the Galactic plane, which is about 30 degrees away from the NCP. 
\section{Results\label{sec:results}}
In this section, we present results mainly to highlight important stages in the calibration and finally, to present the noise limits that we have reached using LOFAR. The results  are based on all three datasets given in Table \ref{tab:results}, unless stated otherwise, and continuum images are made using inverse variance weighted averaging of the 240 subband images.

\subsection{The performance of {\tt SAGECal} in directional calibration}
The effects due to beam shapes and the ionosphere are major causes of errors in LOFAR images. Therefore, directional calibration is essential. We present a few images to highlight the performance of {\tt SAGECal}.

\begin{figure*}[htbp]
\begin{center}
\epsfxsize=5.4in \leavevmode\epsfbox{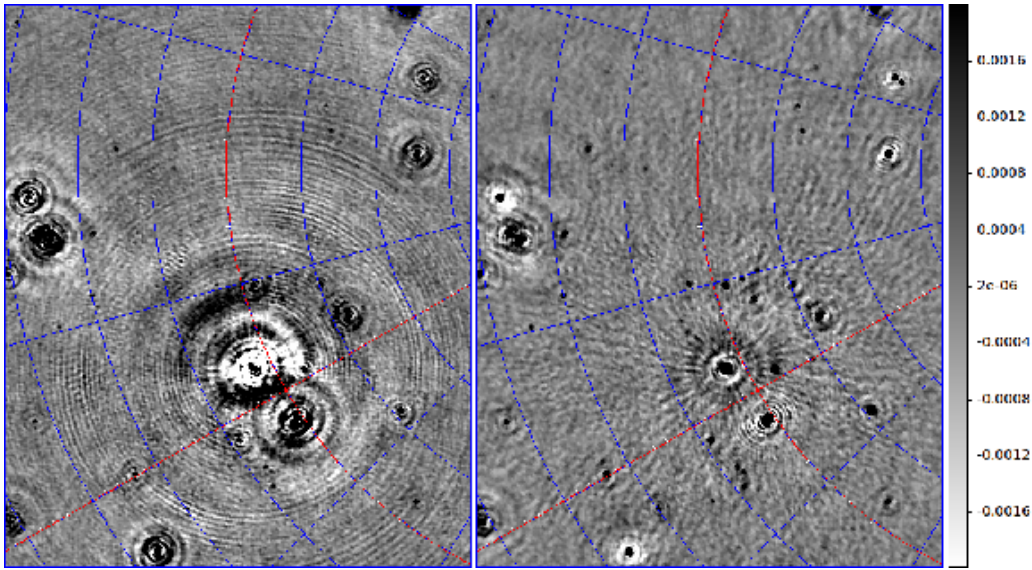}
\caption{An area (about 0.5 deg $\times $ 0.5 deg)  close to the NCP before and after multi-directional calibration using {\tt SAGECal}. The image on the left is before running {\tt SAGECal} and after a deep CLEAN deconvolution and errors due to beam variation and (some) ionospheric variations are clearly visible. On the right hand image, the sources are subtracted directly from the visibility data and the sky model is restored onto the residual image. Most of these errors visible in the left hand image are eliminated in the right hand image as CLEAN based deconvolution fails to consider the directional errors into account. The peak flux is 5 Jy/PSF and the colourbar units are in Jy/PSF.\label{ncp_zoom1}}
\end{center}
\end{figure*}

In Fig. \ref{ncp_zoom1}, we present the area around the brightest source NVSS J011732+892848. The image on the left is before multi-directional calibration and source subtraction, with only a deep CLEAN based deconvolution. The image on the right in Fig.  \ref{ncp_zoom1} is after running {\tt SAGECal} and after restoring the sky model onto the residual image. It is clear that the errors due to beam variations and ionospheric variations have largely been eliminated in the right panel of Fig. \ref{ncp_zoom1}. Some errors are still present, mainly caused by inaccurate source models used in multi-directional calibration. We emphasize that longer baselines are needed to construct accurate models for such complex sources.

\begin{figure*}[htbp]
\begin{center}
\epsfxsize=5.4in \leavevmode\epsfbox{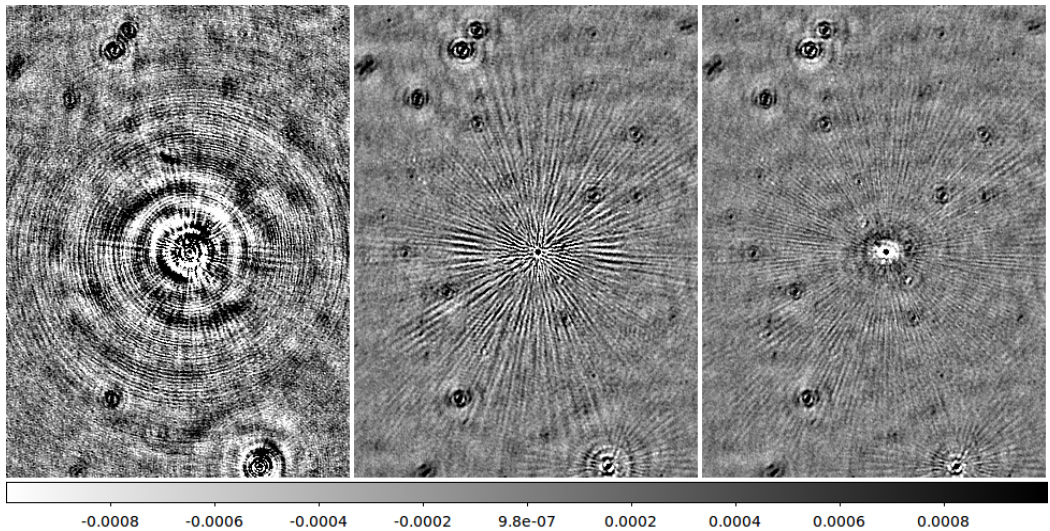}
\caption{The performance of {\tt SAGECal} with different solution intervals. The image on the left is without multi-directional calibration. The image in the middle is after running {\tt SAGECal} with a solution interval of 20 minutes. The image on the right is after running {\tt SAGECal} with a hybrid solution interval, where solutions are obtained along bright source clusters at every 5 minutes and for fainter source clusters, every 20 minutes. It is clear that the small scale ionospheric errors present in the middle figure are mostly eliminated in the right panel. However, ionospheric variations due to decorrelation effects within the 5 minute interval are still present on the right panel. The colourbar (bottom) units are in Jy/PSF.\label{ncp_zoom2}}
\end{center}
\end{figure*}

In Fig. \ref{ncp_zoom2}, we give another example for the effect of the time interval chosen in {\tt SAGECal} for multi-directional calibration. The image on the left is without any multi-directional calibration and significant errors due to beam shape and ionosphere are visible. The image in the middle of Fig. \ref{ncp_zoom2} shows the image obtained after running {\tt SAGECal} with directional calibration performed every 20 minutes. The beam variations, which are slower, are completely eliminated by this procedure. However, the ionospheric variations, that could have time scales much less than 20 minutes are still present.  

The right panel of Fig. \ref{ncp_zoom2} shows the result after running {\tt SAGECal} with a 'hybrid' solution scheme. In this case, we solve for bright sources once every 5 minutes and for fainter sources once every 20 minutes. Most of the ionospheric errors present in the middle panel have been removed in this figure. There are still errors due to inaccurate source models (the source was assumed to be a perfect point source, but this is not accurate enough) and also due to ionospheric phase variations with a time scale smaller than 5 minutes.

The time interval of 20 minutes chosen for obtaining the solutions gives us 120 time samples (each sample is of 10 s duration). For each time sample, we have 990 baselines with 45 stations. Therefore, we have about 8$\times$120$\times$990$=$1 million real constraints to obtain a solution. The number of real parameters in a solution is 45$\times$150$\times$8$=$54000 for 150 directions in the sky. Therefore, the ratio between the number of constraints and the number of parameters is about 18 which is more than sufficient to obtain a reliable solution. This can be further improved by using data points at different frequencies as proposed by \cite{Bregman}.
\subsection{Widefield images}
\begin{figure*}[htbp]
\begin{center}
\epsfxsize=7.4in \leavevmode\epsfbox{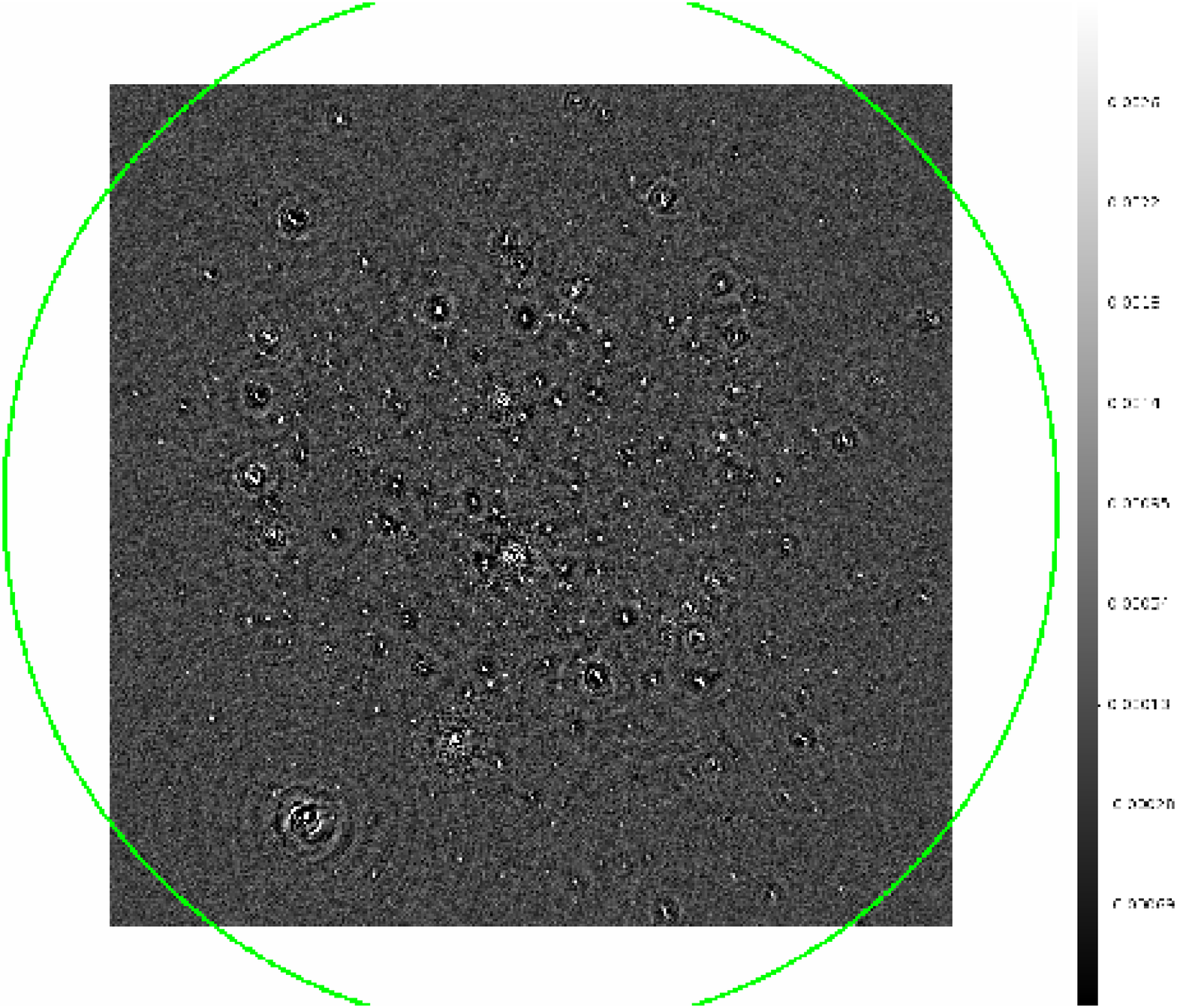}
\caption{The NCP image after calibration, but before running {\tt SAGECal}, which has also been deconvolved using CASA. The circle indicates an area of diameter 10 degrees. The source 3C61.1 is at the bottom left hand corner. The image has $7200\times 7200$ pixels of size 4$^{\prime\prime}$ with a PSF of 12$^{\prime\prime}$ and the noise level at this stage is still about 400 $\mu$Jy/PSF. The colourbar units are in Jy/PSF.\label{ncp_calib_fov}}
\end{center}
\end{figure*}

We present widefield  images obtained for the full dataset given in Table \ref{tab:results}. First, in Fig. \ref{ncp_calib_fov}, we present the image obtained after calibration as described in section \ref{calib_corr}, but before running {\tt SAGECal}. Therefore, no source subtraction is performed and only traditional CLEAN based deconvolution has been applied. The circle indicates an area of diameter 10 degrees. The peak flux of this image is about 5.3 Jy and the noise level is about 400$\mu$Jy/PSF. The complex source 3C61.1 is at the bottom left hand corner. The striking features in this image are the artefacts surrounding almost every source. As described previously, there are three major reasons for these artefacts: (i) varying LOFAR beam shapes which are different for each station,  (ii) ionospheric phase errors, and (iii) classical deconvolution errors due to having partially resolved sources. For instance for the case of 3C61.1, all three of the aforementioned causes create artefacts, which are clearly visible close to the bottom left hand corner.

\begin{figure*}[htbp]
\begin{center}
\epsfxsize=7.4in \leavevmode\epsfbox{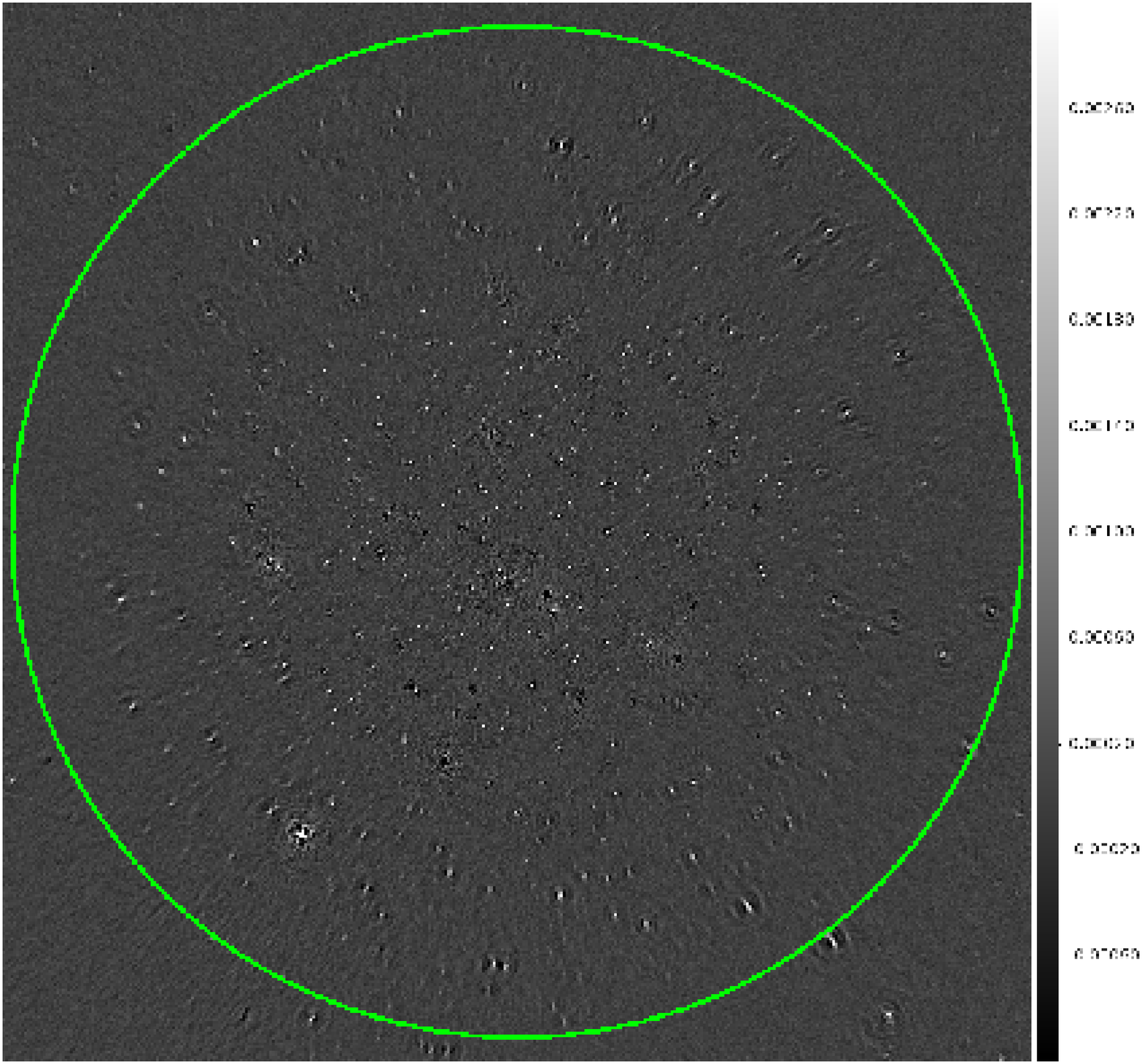}
\caption{The NCP image after multi-directional calibration and source subtraction using {\tt SAGECal}. After a shallow deconvolution using CASA (mainly to estimate the PSF), the skymodel is restored onto the image. The  circle indicates an area of diameter 10 degrees. The image has $12000\times12000$ pixels of size 4$^{\prime\prime}$ with a PSF of 12$^{\prime\prime}$ and the noise level is about 100 $\mu$Jy/PSF. Due to the fact that RS and CS beam shapes have different FOVs the sources at the edge of the image are distorted. In addition, due to frequency smearing, the sources at the edge of the image appear 'attracted' towards the center. The colourbar units are in Jy/PSF.\label{ncp_calib_sage}}
\end{center}
\end{figure*}

The only way to improve the image in Fig. \ref{ncp_calib_fov} is multi-directional calibration as described in section \ref{calib_sub}. We have shown the image obtained after running {\tt SAGECal} in Fig. \ref{ncp_calib_sage}. The circle indicates an area of diameter 10 degrees. Comparison of Figs. \ref{ncp_calib_fov} and  \ref{ncp_calib_sage} shows that most significant artefacts in Fig. \ref{ncp_calib_fov} have been eliminated in Fig. \ref{ncp_calib_sage}. The prominent artefacts that still remain are due to the fact that CS and RS beam shapes have different FOVs and also due to frequency smearing.  We now reach a noise level of about 100 $\mu$Jy/PSF at the outskirts of Fig.  \ref{ncp_calib_sage}, while the peak value in the image is about 5.3 Jy. This corresponds to a formal dynamic range of 50,000:1. 

\begin{figure*}[htbp]
\begin{center}
\epsfxsize=7.4in \leavevmode\epsfbox{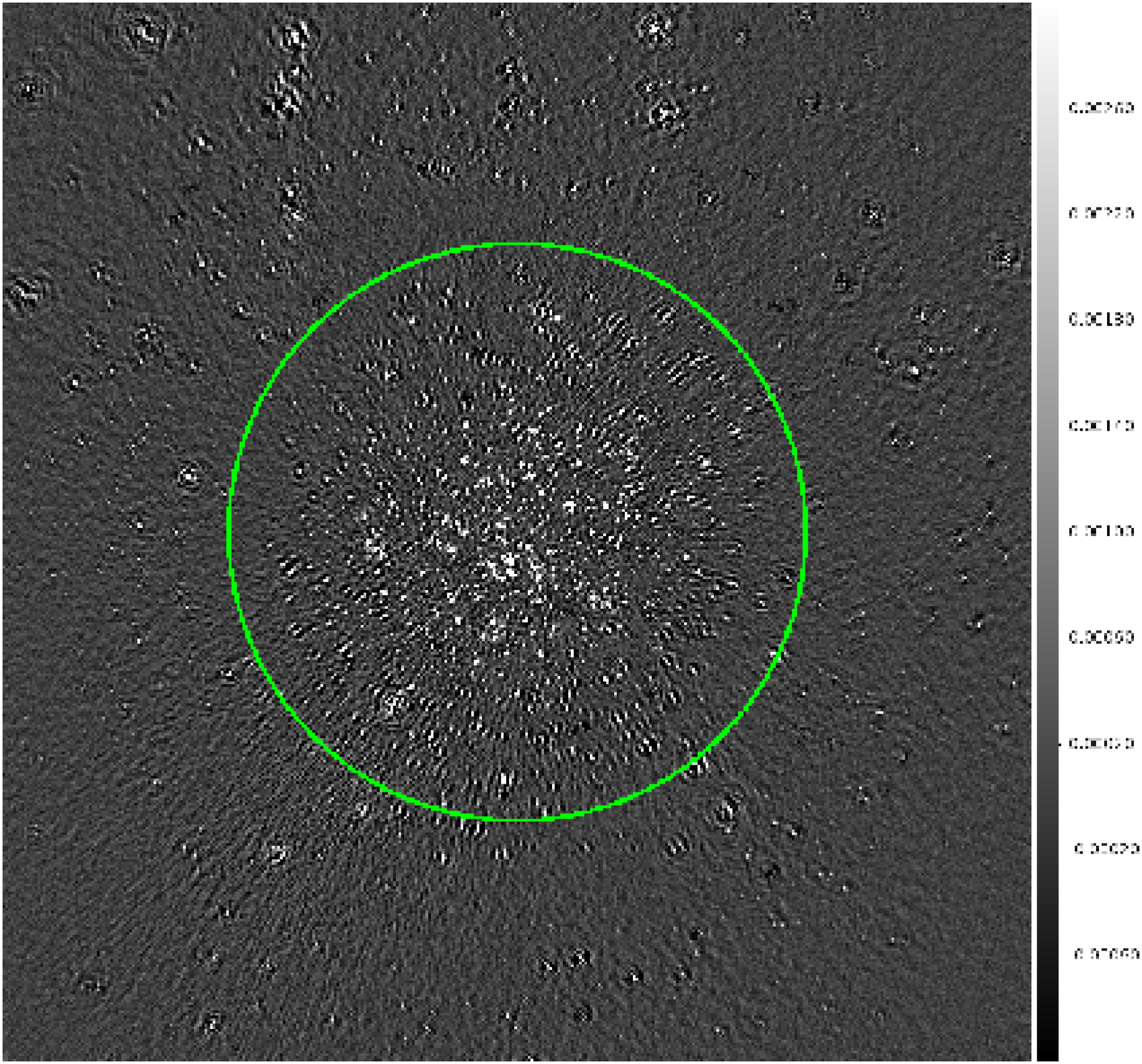}
\caption{The NCP image after multi-directional calibration and source subtraction using {\tt SAGECal}, using only the short baselines. After a shallow deconvolution using CASA (mainly to estimate the PSF), the sky model is restored onto the image. The  circle indicates an area of diameter 10 degrees. The image has $2000\times2000$ pixels of size 35$^{\prime\prime}$ with a PSF of 150$^{\prime\prime}$ and the noise is about 300 $\mu$Jy/PSF. The colourbar units are in Jy/PSF.\label{ncp_calib_sage_core}}
\end{center}
\end{figure*}

In Fig. \ref{ncp_calib_sage_core}, we give the image made only with the short baselines ($< 1200$ wavelengths) using the same data of Fig. \ref{ncp_calib_sage}. In this image, the circle shows an area with 10 degrees in diameter and the density of the sources close to this circle  is clearly less than in other areas of the image. We also see a significant number of sources away from the FOV that are seen through sidelobes of the beam. Most of these sources have not been included in our multi-directional calibration and hence, they have significant artefacts. The noise level in this image is about 300 $\mu$Jy/PSF with the peak flux of about 5.3 Jy. Note, however, that the noise level is a strong function of the distance from the field centre. 

\subsection{New sources}
\begin{figure*}[htbp]
\begin{center}
\begin{minipage}{1.00\linewidth}
\begin{minipage}{0.50\linewidth}
\begin{center}
\epsfxsize=2.4in \leavevmode\epsfbox{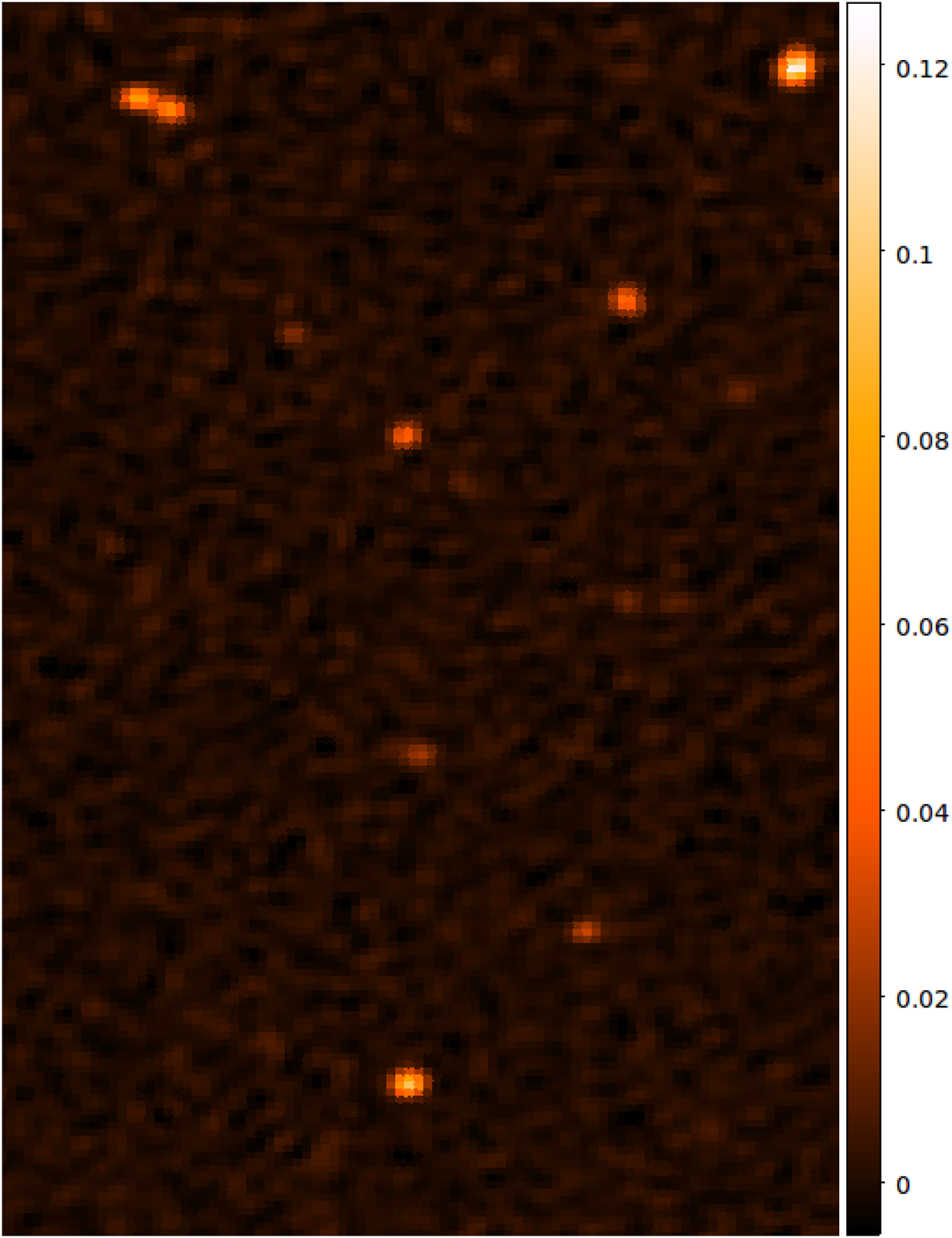}
\end{center}
\end{minipage}
\begin{minipage}{0.50\linewidth}
\begin{center}
\epsfxsize=2.4in \leavevmode\epsfbox{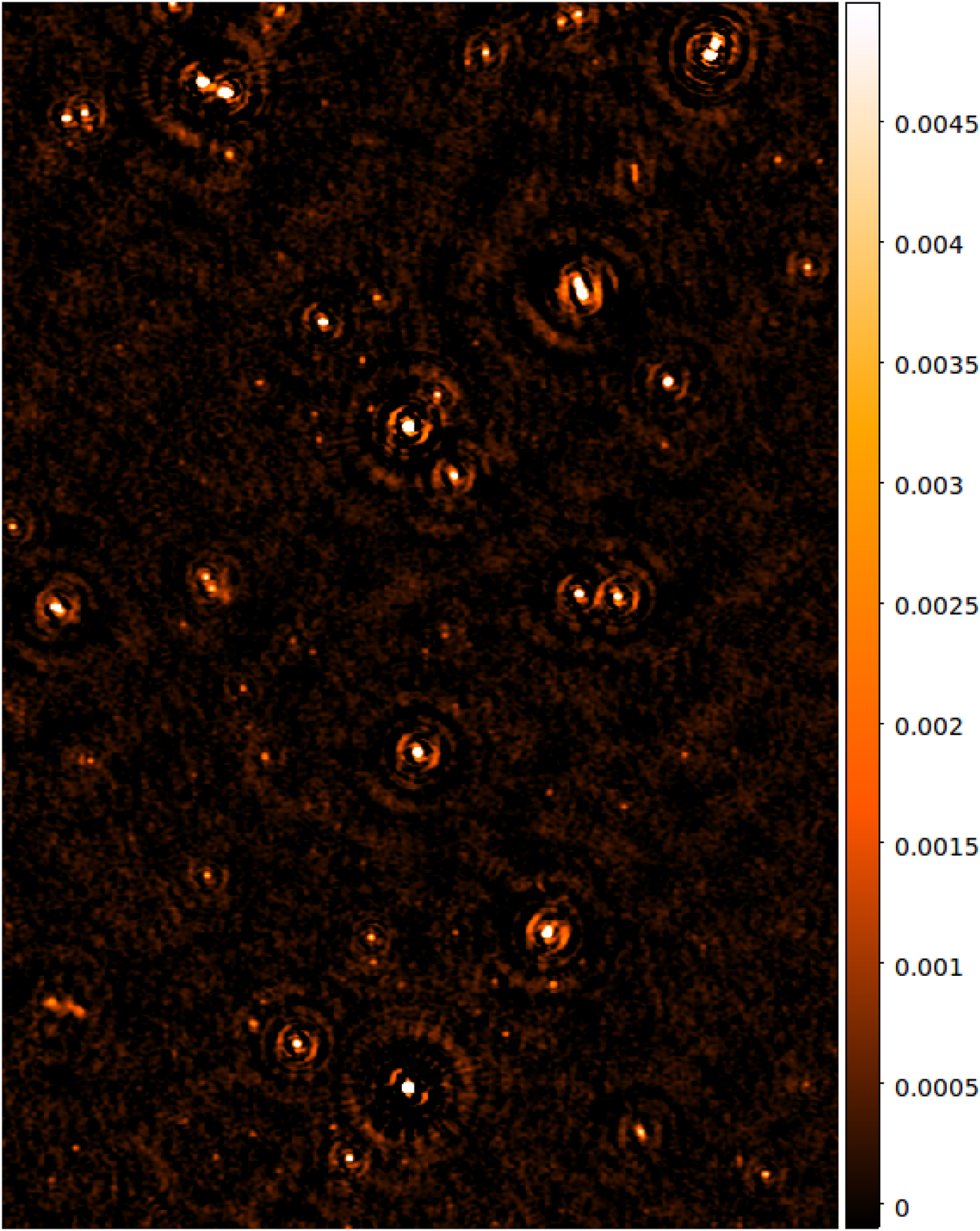}
\end{center}
\end{minipage}
\caption{Comparison of a small area of the NCP image, size $0.6\times1.0$ degrees with WENSS. The left panel shows the image from WENSS (PSF 60$^{\prime\prime}$) while the right panel shows the equivalent image made using LOFAR (PSF 12$^{\prime\prime}$) after running {\tt SAGECal} and a shallow deconvolution. The colourbar units are in Jy/PSF. Many more sources, at much higher angular resolution can be detected. \label{ncp_wenss_vlss}}
\end{minipage}
\end{center}
\end{figure*}

Since we reach a noise limit of about 100 $\mu$Jy/PSF, we detect a large number of sources that have not been detected in previous  observations, even at higher frequencies. In Fig. \ref{ncp_wenss_vlss}, we present a small area (0.6$\times$1.0 degrees) of Fig. \ref{ncp_calib_sage} to compare  to an image from WENSS.

\begin{figure*}[htbp]
\begin{center}
\begin{minipage}{1.00\linewidth}
\begin{minipage}{0.50\linewidth}
\begin{center}
\epsfxsize=3.2in \leavevmode\epsfbox{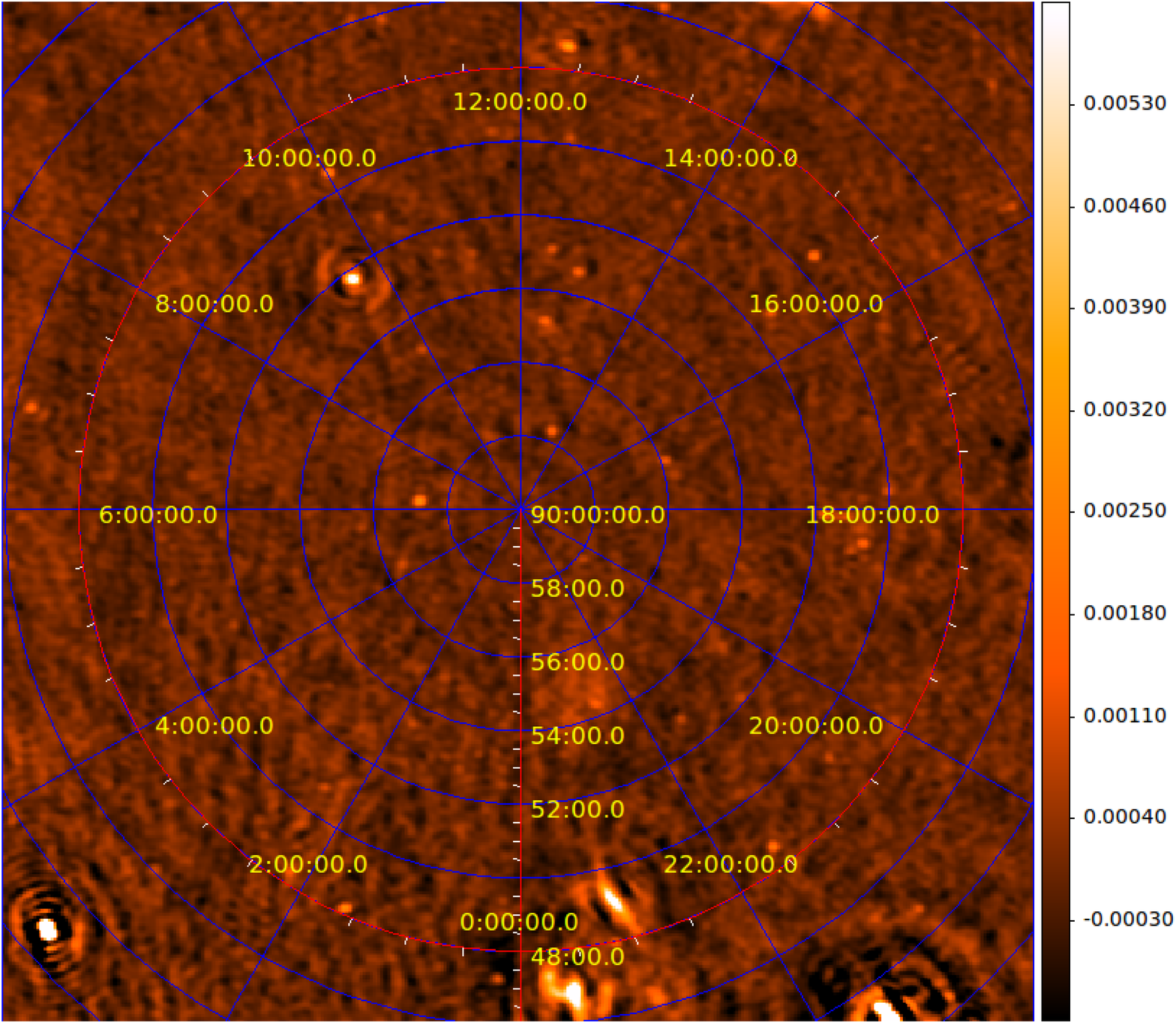}
%\vspace{0.1cm}\centerline{(a)}
\end{center}
\end{minipage}
\begin{minipage}{0.50\linewidth}
\begin{center}
\epsfxsize=3.2in \leavevmode\epsfbox{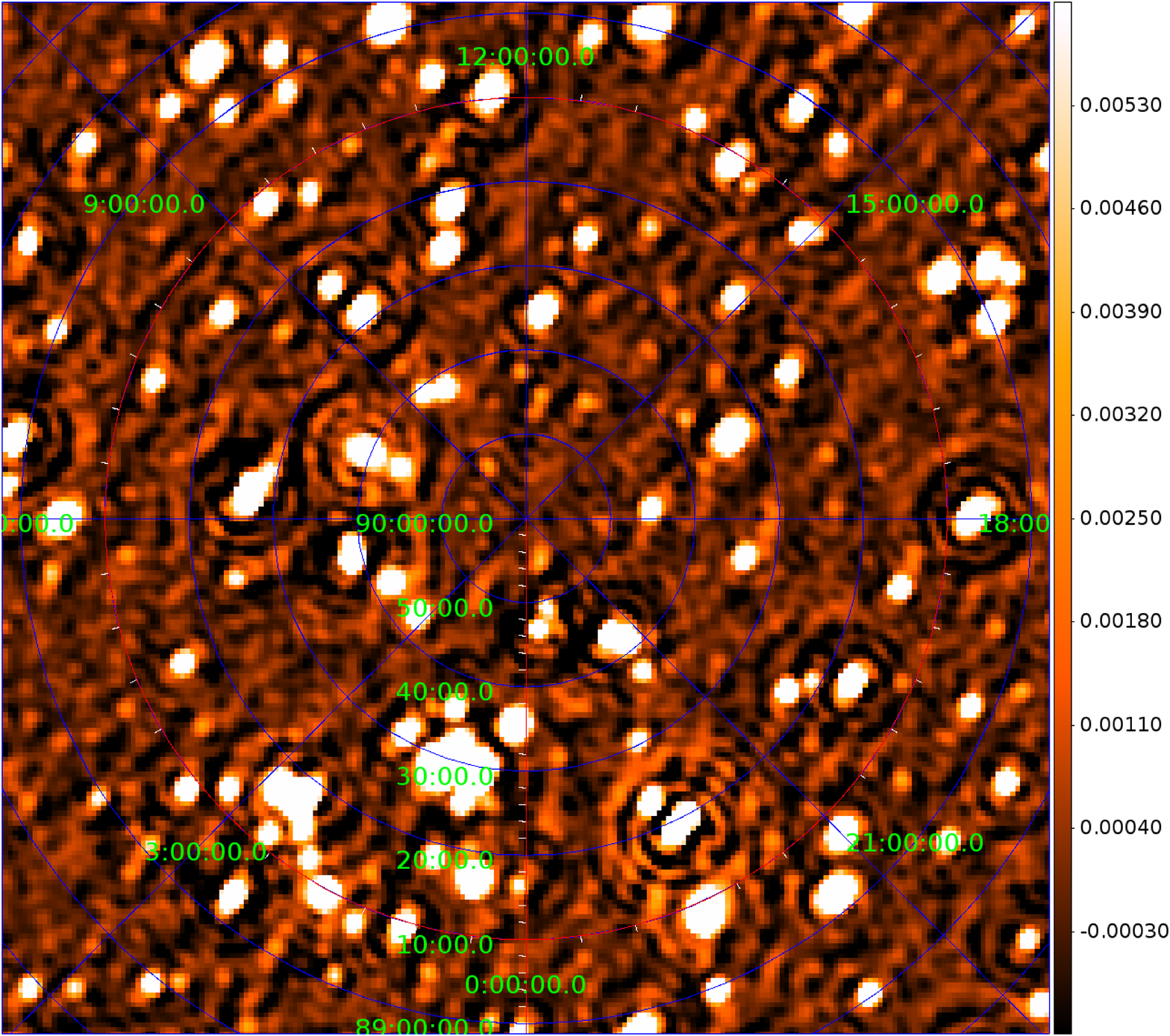}
%\vspace{0.1cm}\centerline{(b)}
\end{center}
\end{minipage}
\end{minipage}
\caption{Images of the area close to the NCP. The image on the left panel is using all baselines and has a pixel size of 4${^{\prime\prime}}$ and a PSF of 12${^{\prime\prime}}$. The image on the right is using core only baselines with a pixel size of 35${^{\prime\prime}}$ and a PSF of 150${^{\prime\prime}}$. The colourbar units are in Jy/PSF.\label{ncp_zoom3}}
\end{center}
\end{figure*}

We present an area close to the NCP in Fig. \ref{ncp_zoom3}.  The left panel in Fig. \ref{ncp_zoom3} shows an image made with all baselines which gives a PSF of 12${^{\prime\prime}}$. The right panel in Fig. \ref{ncp_zoom3} shows an image made using only the core baselines and has a PSF of about 150${^{\prime\prime}}$. An important lesson that can be drawn from Fig. \ref{ncp_zoom3} is the total absence of artefacts close to the pole. If there would be any residual geostationary RFI, their effects would accumulate near the pole. However, we see no unexplained artefacts.

\subsection{Effects of bright sources at large angular distances}
\begin{figure*}[htbp]
\begin{center}
\epsfxsize=6.4in \leavevmode\epsfbox{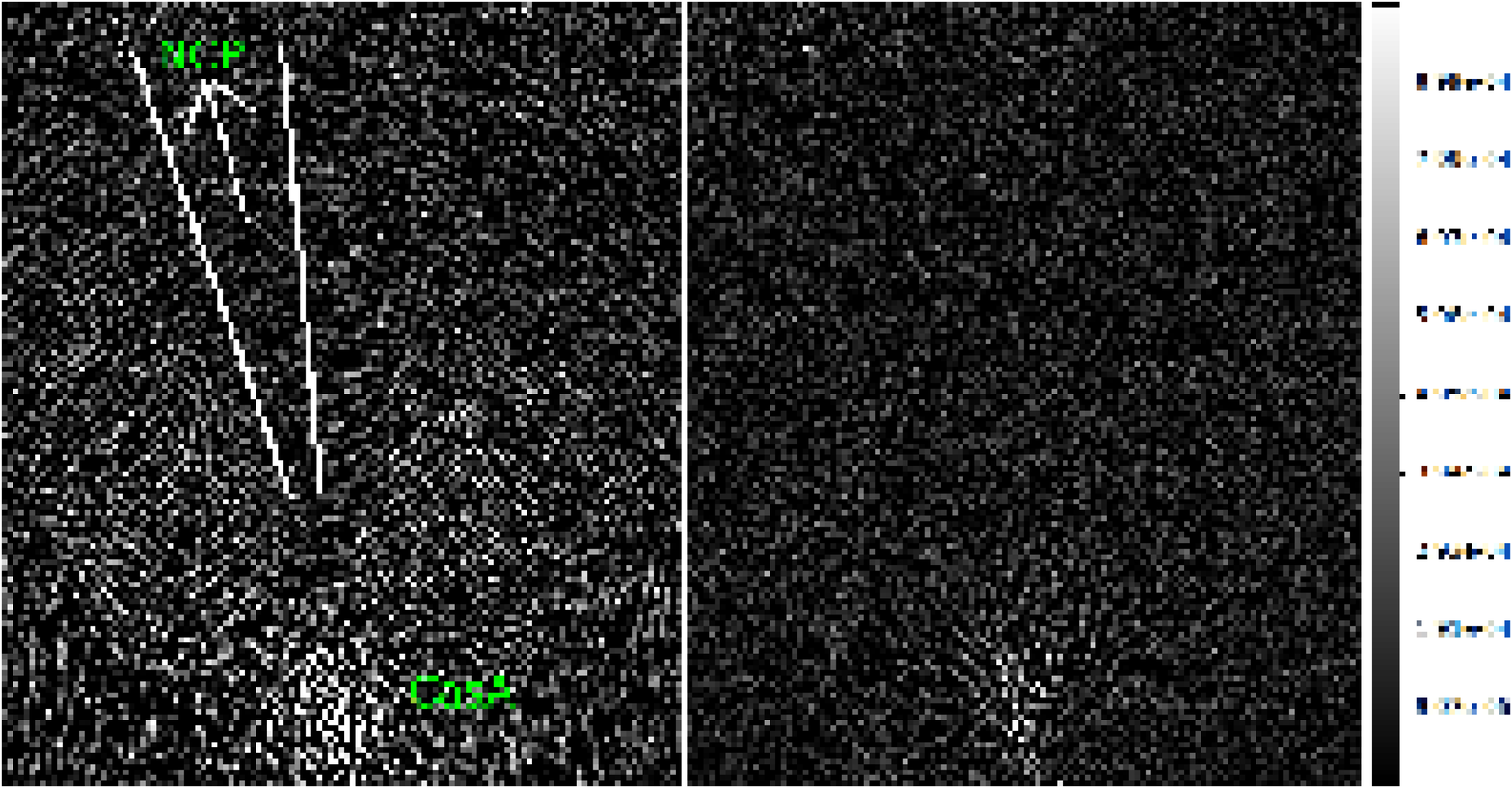}
\caption{Images of CasA, which is about 30 degrees away from the NCP, made only using the short baselines ($< 1200$ wavelengths). The distorted image of CasA (due to smearing and directional errors) is at the bottom of these figures. The image on the left is obtained after running {\tt SAGECal} without taking CasA into account. The image on the right is after adding CasA to the sky model and running {\tt SAGECal}. It is clear that the ripples on the left panel are eliminated on the right panel. Moreover, the left panel shows a 'cone' directed towards the NCP where the ripples are absent. The colourbar units are in Jy/PSF.\label{ncp_casa}}
\end{center}
\end{figure*}

As shown in Fig. \ref{FOV}, there are a few bright sources in the neighborhood of the NCP. We have already mentioned 3C61.1 which is still well inside the FOV. The other source that has a significant  effect is CasA, which is about 30 degrees away from the NCP. In Fig. \ref{ncp_casa}, we present the images around CasA, made with only the core station baselines. The images with baselines using core stations only are more affected by CasA than images that include remote stations. There are at least four reasons that contribute to this. First, core stations have wider station beams (compared to a remote stations) and therefore, CasA is less attenuated on core-core baselines . Secondly, the core stations have more short baselines than remote stations, hence see more flux from CasA, which is heavily resolved at baselines longer than 1000$\lambda$. Thirdly,  time and frequency smearing lead to a significant attenuation of the visibilities of distant sources. Fourthly, ionospheric effects such as non-isoplanaticity rapidly increase with the length of the interferometer baseline.   

In Fig. \ref{ncp_casa} left panel, we show an image where we have run {\tt SAGECal} while ignoring the effect of CasA. In other words, we did not include CasA in our sky model and neither did we solve along the direction of CasA. CasA is at the bottom of this image and is heavily distorted due to beam and ionospheric errors as well as time and frequency smearing. The ringlike structures centered at CasA and spreading throughout the image is clearly visible. However, there is an area shaped like a cone that is directed towards the NCP where there are no ripples. This is due to the fact that multi-directional calibration that ignores CasA (mainly close to the NCP) has absorbed the effect of CasA directed towards the NCP. Similar effects can be seen in sequential source subtraction schemes such as 'peeling'.

In the right panel of Fig. \ref{ncp_casa}, we have included CasA in our multi-directional calibration using {\tt SAGECal}. Most of the ripples in the left panel are eliminated in the right panel of Fig. \ref{ncp_casa}. Furthermore, the noise level is reduced by a few percent after including CasA in the calibration. There are still some errors close to the location of CasA. This is mainly due to errors in the CasA source model used in the subtraction and we expect to get better results with an updated source model \citep{CasASBY}.

Because CasA is a strong source, we clearly see its effect as we have just described. Conversely, even faint sources would have a similar effect, albeit at a low magnitude. We perform a statistical analysis of this effect due to faint sources in section \ref{sec:outlier}.

\subsection{Noise}
\begin{figure}[htbp]
\begin{center}
\epsfxsize=3.4in \leavevmode\epsfbox{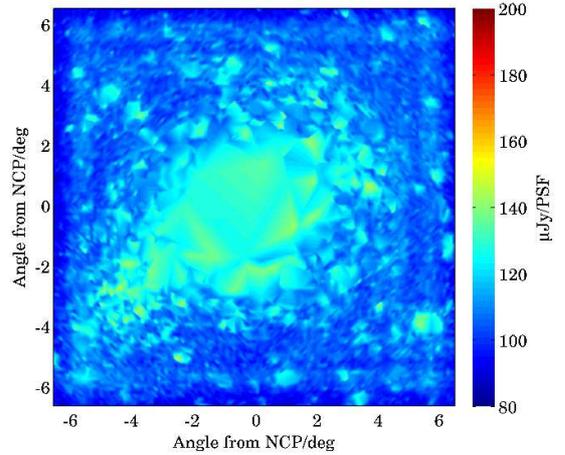}
\caption{Variation of noise across the image shown in Fig. \ref{ncp_calib_sage}. The noise is estimated using a moving rectangular window of about $400\times400$ pixels and the noise is only calculated when there are no sources ($>$ 0.5 mJy) within the rectangular window. The noise variation is shown across a 13$\times$13 degrees FOV.\label{noise_map}}
\end{center}
\end{figure}

One important question that needs to be answered is whether we have reached the theoretical noise limits in our images and if not, provide plausible explanations for the difference. The theoretical noise  can be calculated as follows \citep{LOFARsens}:
\beq \label{thnoise}
\rm{noise}_{\rm{subband}} =  \frac{w_{\rm{imaging}}}{\sqrt{4 \Delta_f \Delta_t  \left(\frac{N_c(N_c-1)}{2S_c^2}+\frac{N_cN_r}{S_c S_r} + \frac{N_r(N_r-1)}{2S_r^2}\right)}}.
\eeq 
In (\ref{thnoise}), the noise per subband (bandwidth 183 kHz) is given by $\rm{noise}_{\rm{subband}}$ while the system equivalent flux densities for core and remote stations are given by $S_c$ and $S_r$, respectively. We assume $S_c=3360$ Jy and $S_r=1680$ Jy, taking the proximity to the Galactic plane into account. The total bandwidth and integration time are considered as $\Delta_f$=183 kHz, and $\Delta_t$=6$\times$3600 s. The total number of core and remote stations are given by $N_c$ and $N_r$. The scale factor $w_{\rm{imaging}}$ is used to take imaging weights into account and we take this equal to $2$ for uniform weighting. For the observation with highest sensitivity from Table \ref{tab:results} (L24560), with $N_c=38$, $N_r=7$, we get $\rm{noise}_{\rm{subband}}=1.46$ mJy. Moreover, for the full observation, with $229$ subbands, and taking the variation of $S_c$ and $S_r$ with frequency into account, we get a theoretical noise value of about 93 $\mu$Jy. As seen in Table \ref{tab:results}, we are only a factor of 1.4 from reaching the theoretical noise limit. An important question is how to estimate the noise in the images, in other words, which area of the image to use to calculate the noise. In Fig. \ref{noise_map}, we show the variation of the noise across the 13$\times$13 degrees image shown in Fig. \ref{ncp_calib_sage}. As expected, the noise close to the NCP is higher and is about 180 $\mu$Jy due to additional contamination by unsubtracted compact sources as well as by diffuse foregrounds. However, with the subtraction of many more sources as well as foregrounds, we expect to reach the noise that we see at the edge of the image in Fig. \ref{noise_map}. Therefore, we study the behavior of the noise seen at the edge of this image as this is what we intend to reach. We emphasize  that the achieved continuum noise levels in low resolution images, i.e. made using only the core stations,  will be limited by classical confusion noise in the inner parts of the images. \citet{Ber1} and \citet{Pizzo} estimated this to be about 0.6 mJy for the WSRT at a frequency of 140~MHz.  Because the LOFAR core has a similar extent as the WSRT we expect this to be the asymptotic limit of the continuum noise level in core-only LOFAR observations.    

\begin{figure}[htbp]
\begin{center}
\epsfxsize=3.4in \leavevmode\epsfbox{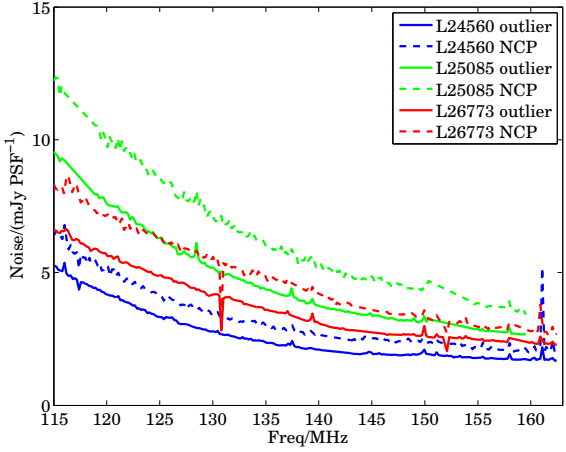}
\caption{Noise in images made using three nights of data. The noise of each image is estimated at a location at the edge of the FOV as well as at the NCP, using a rectangular window of about $400\times400$ pixels. The three different colours correspond to the three different nights listed in Table \ref{tab:results}.\label{ncp_noise_abs}}
\end{center}
\end{figure}

To study the noise behavior in more detail, we provide several figures where we give the noise (or the standard deviation) of a small rectangular area in the images (about 4 degrees away from the NCP) at different frequencies and using different observations listed in Table  \ref{tab:results}. In Fig. \ref{ncp_noise_abs}, we present noise levels per subband (standard deviation), determined with images made using the three nights with imaging parameters as in Fig. \ref{ncp_calib_sage}. Additionally, we also show the noise estimated at the NCP in Fig. \ref{ncp_calib_sage}.
The main conclusion that can be drawn from Fig.  \ref{ncp_calib_sage} is the variability of the noise (or the sensitivity) of LOFAR from night to night. This is due to some stations not working or not working at full sensitivity at different nights. We expect this to be much more stable before the commencement of dedicated EoR observations.

The best night in Fig. \ref{ncp_noise_abs} is for the observation number L24560 of Table \ref{tab:results} which has a noise level of about 2 mJy per subband at the high frequency end. There is a steep rise in the noise level at frequencies below 130 MHz. This we attribute to the rising  contribution of Galactic background noise, increased sidelobe noise from an increasing number of faint background sources (due to both a wider primary beam and steeper source spectra), emission from  the Galactic plane and other very bright  sources like CygA (see section \ref{sec:outlier}). There are also some spikes and dips in the noise curves where RFI removal has flagged significant amounts (more than 30\%) of data. We have discarded such images from further analysis.

\begin{figure}[htbp]
\begin{center}
\epsfxsize=3.4in \leavevmode\epsfbox{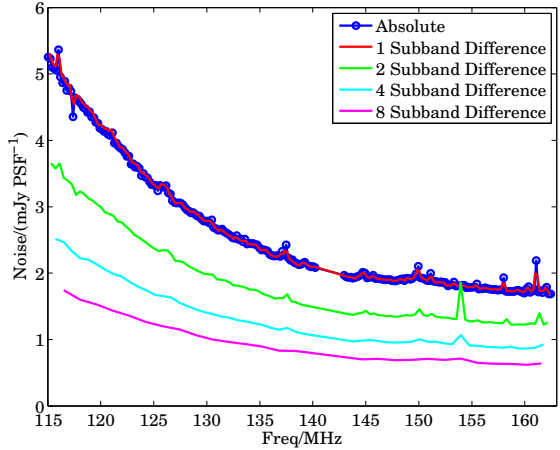}
\caption{Image difference noise of images made using one night (L24560) of data. The noise of each image is estimated  using a rectangular window of about $400\times400$ pixels, after taking the difference of images adjacent in frequency. We have also averaged images adjacent in frequency into groups of 2,4 and 8 and have taken their difference as well.\label{ncp_noise_diff}}
\end{center}
\end{figure}

In Fig. \ref{ncp_noise_diff}, we give the image difference noise estimates for the best night (L24560) of Table \ref{tab:results}. The image difference noise is calculated by (i) subtracting two images adjacent in frequency (ii) estimating the standard deviation of the subtracted image (iii) multiplying the standard deviation by $1/\sqrt{2}$. We have also plotted the absolute noise level in the same figure. It is clear from Fig.  \ref{ncp_noise_diff} that the absolute and differential noise curves are almost identical. This suggests that our absolute noise estimate does not contain any frequency independent systematic effects, except systematic effects that are uncorrelated between adjacent images. Future processing would take the difference at narrower frequencies to verify this.

We have also plotted in Fig. \ref{ncp_noise_diff} image difference noise curves where instead of taking the difference of adjacent subbands, we have averaged $B$ images (or subbands) together and calculated the $B$ difference between images. We have done this for $B=2,4,8$ in Fig.  \ref{ncp_noise_diff}. As $B$ increases, the differential noise should be more affected by systematic effects due to the large bandwidth of averaging. This, in turn should be reflected by the differential noise not decreasing as $1/\sqrt{B}$. However, we do see a decrease by $1/\sqrt{B}$ in the plots in  Fig. \ref{ncp_noise_diff}, even when $B=8$.

\begin{figure}[htbp]
\begin{center}
\epsfxsize=3.4in \leavevmode\epsfbox{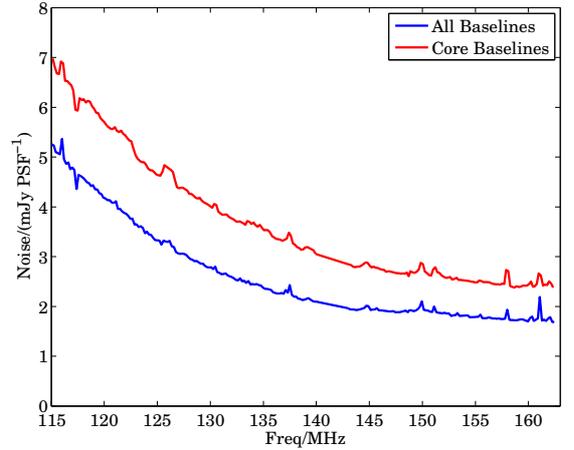}
\caption{Comparison of noise in images made with all baselines (blue) and core only ($< 1200$ wavelengths) baselines (red).\label{ncp_noise_coreabs}}
\end{center}
\end{figure}

We also show the noise comparison between the images made with all baselines and with the images made with core only ($< 1200$ wavelengths) baselines in Fig. \ref{ncp_noise_coreabs}. The corresponding continuum image is given in Fig. \ref{ncp_calib_sage_core} where the imaging parameters are also given. We have estimated the noise of image with core only baselines at a location about 5 degrees away from the center of Fig. \ref{ncp_calib_sage_core}.

\begin{figure}[htbp]
\begin{center}
\epsfxsize=3.4in \leavevmode\epsfbox{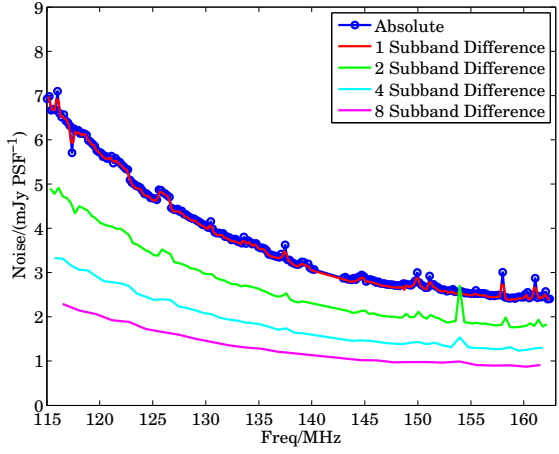}
\caption{Image difference noise of images made with core only ($< 1200$ wavelengths) baselines for one night (L24560) of data. We have also plotted the absolute noise which agrees well with the differential noise.  We have also averaged images adjacent in frequency  into groups of 2,4 and 8 and have taken their difference as well.\label{ncp_noise_corediff}}
\end{center}
\end{figure}

In Fig. \ref{ncp_noise_corediff}, we give the image difference noise plots for images made only with core baselines. We have also plotted the absolute noise curve which agrees well with the image difference noise. Similar to Fig. \ref{ncp_noise_diff}, we have also averaged adjacent subbands of groups 2,4 and 8 and found the differential noise between averaged images.

\begin{figure}[htbp]
\begin{center}
\epsfxsize=3.4in \leavevmode\epsfbox{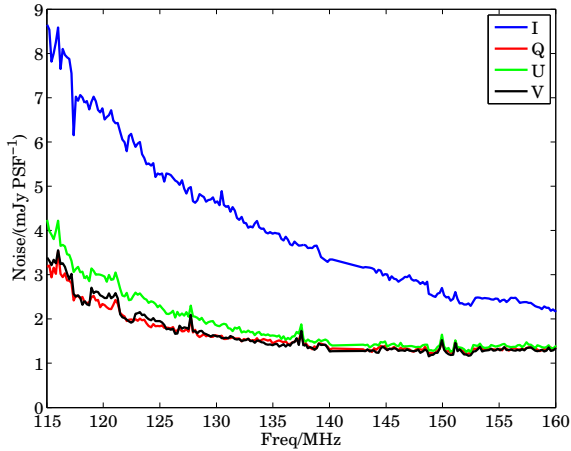}
\caption{Noise in all four Stokes images for one night (L24560) of observed data. The images are made with core only ($< 1200$ wavelengths) baselines. At the high frequency end, the noise in total intensity is about 1.6 times higher than the noise in polarization.\label{ncp_noise_pol}}
\end{center}
\end{figure}

In Fig. \ref{ncp_noise_pol}, we show the noise plots for all 4 Stokes images (I,Q,U,V) for one night (L24560) of observed data. The noise level of Stokes I is significantly higher which we attribute to unsubtracted sources. We give a detailed analysis of this in section \ref{sec:outlier}. 

\subsection{Linear polarization}
Due to the fact that we correct for the element (dipole) beam polarization along the direction of the NCP during calibration, we do not see substantial instrumental polarization in our Stokes $Q$,$U$ and $V$ images. Note also that we expect most of the sources to be intrinsically unpolarized in this frequency range. Even though we only correct for the element beam gain along the direction of the NCP, the relative variation within the full FOV of the dipole beam shape is very little (therefore this correction is satisfactory within the full FOV). However, we expect to see an increase in instrumental polarization at the edge of the FOV but the station beam attenuation makes this instrumental polarization less obvious. During the multi-directional calibration phase using {\tt SAGECal}, we also solve for a full Jones matrix and therefore, what remains of this instrumental polarization is correctly subtracted.

\begin{figure*}[htbp]
\begin{center}
\epsfxsize=5.4in \leavevmode\epsfbox{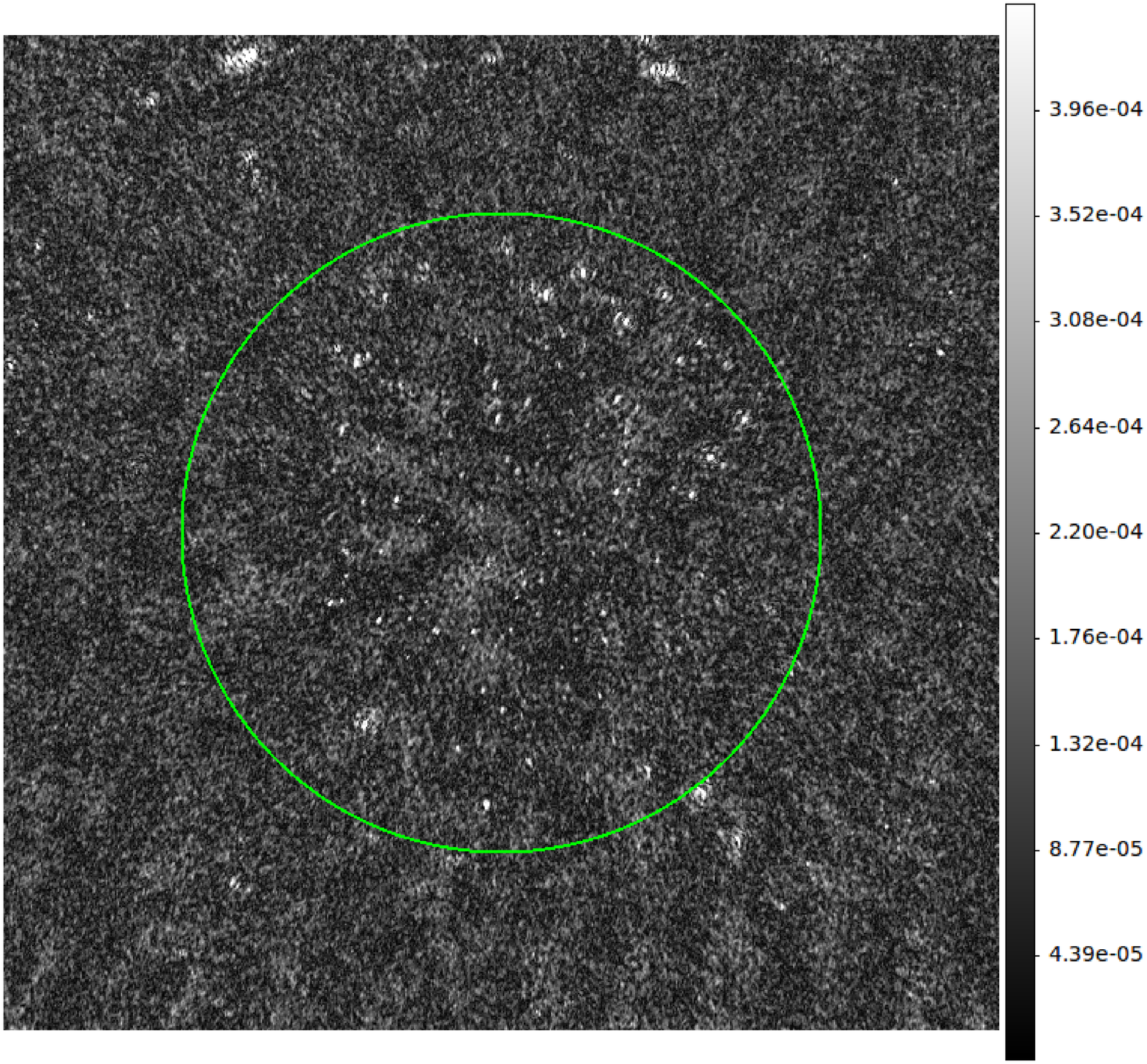}
\caption{Total polarized intensity image at RM=0~rad m$^{-2}$ using core only baselines (150$^{\prime\prime}$ PSF), after running {\tt SAGECal}. The circle indicates an area of 10 degrees in diameter. The noise is about 110 $\mu$Jy/PSF. The colourbar units are in Jy/PSF.\label{rm0_pol}}
\end{center}
\end{figure*}

In order to detect any diffuse Galactic foregrounds that might appear in polarized images, we have performed rotation measure (RM) synthesis \citep{RMB} using data from one of the nights listed in Table \ref{tab:results} (L24560). We show the total polarized intensity image for RM=0~rad m$^{-2}$ in Fig. \ref{rm0_pol}. The noise in this image is about 110 $\mu$Jy/PSF and apart from many weakly instrumentally polarized discrete sources, we detect only very faint diffuse structure, at any value of RM. However, we have seen that subtraction of compact sources using {\tt SAGECal} would suppress the unmodeled diffuse foregrounds. To alleviate this from happening in future processing, we would ignore the contribution from short baselines while running {\tt SAGECal} for calibration along multiple directions.

\section{Effect of Outlier Sources in Image Difference Noise\label{sec:outlier}}
In this section, we present an analysis of the contribution of sources far away from the field center to the (differential) noise of images made at the NCP. In fact, this analysis can be extended to any interferometric observation. We show that our ignorance of these sources indeed act as an additional source of noise.

Consider an elementary interferometer. The visibility $V(u_p,v_p,w_p)$ at coordinates $u_p,v_p,w_p$ on the $uv$ plane is
\beqn \label{vis} \nonumber
\lefteqn{V(u_p,v_p,w_p)=}&&\\
&&\int\int S(l,m) e^{-j 2\pi \frac{f}{c}\left(u_p l+v_p m+ w_p(\sqrt{1-l^2-m^2}-1)\right)} \frac{dl dm}{\sqrt{1-l^2-m^2}}
\eeqn
where $S(l,m)$ is the sky flux density and $l,m$ are the direction cosines. The frequency of the observation is $f$ while the speed of light is $c$.
We assume the sky to consist of a set of discrete sources,
and we arrive at
\beq \label{visdt}
V(u_p,v_p,w_p)=\sum_q I(l_q,m_q) e^{-j 2\pi \frac{f}{c}\left(u_p l_q+v_p m_q+ w_p(\sqrt{1-l_q^2-m_q^2}-1)\right)}
\eeq
where $I(l,m)$ is the sky intensity.

Considering a bandwidth of $\Delta$ for smearing, we have the smeared value of $V(u_p,v_p,w_p)$ around frequency $f_0$ as
\beq
\bar{V}(u_p,v_p,w_p)=\frac{1}{\Delta} \int_{f_0-\Delta/2}^{f_0+\Delta/2} V(u_p,v_p,w_p) df
\eeq
and assuming $I(l_q,m_q)$ variation is small over this bandwidth, this reduces (\ref{visdt}) to 
\beqn \label{visdt_smear}
\lefteqn{\bar{V}(u_p,v_p,w_p)=}&&\\
&&\sum_q I(l_q,m_q) e^{-j 2\pi \frac{f}{c}\left(u_p l_q+v_p m_q+ w_p(\sqrt{1-l_q^2-m_q^2}-1)\right)}\\\nonumber
&&\times \sinc\left( \pi\frac{\Delta}{c} \left(u_p l_q+v_p m_q+ w_p(\sqrt{1-l_q^2-m_q^2}-1)\right) \right).
\eeqn

Let $M$ denote the number of samples in the $uv$ plane and $N$ denote the number of sources in the sky. We represent the visibilities at all points on the $uv$ plane in vector form as $\widetilde{\bf b}$ (size $M\times 1$) and the intensities of all discrete sources in the sky as ${\bf b}$ (size $N\times 1$) where
\beq
\widetilde{\bf b}=\left[ \begin{array}{c}
\bar{V}(u_1,v_1,w_1)\\
\bar{V}(u_2,v_2,w_2)\\
\ldots\\
\bar{V}(u_M,v_M,w_M)
\end{array} \right],\ \ 
{\bf b}=\left[ \begin{array}{c}
I(l_1,m_1)\\
I(l_2,m_2)\\
\ldots\\
I(l_N,m_N)
\end{array} \right].
\eeq

We can relate $\widetilde{\bf b}$ and ${\bf b}$ as
\beq \label{Tf}
\widetilde{\bf b} = {\bf T}_{f_0} {\bf b}
\eeq
where the elements in the matrix ${\bf T}_{f_0}$ (size $M\times N$) are given as
\beqn \label{Tfelem}
\lefteqn{[{\bf T}_{f_0}]_{pq}=}&&\\\nonumber
&&e^{-j 2\pi \frac{f_0}{c}\left(u_p l_q+v_p m_q+ w_p(\sqrt{1-l_q^2-m_q^2}-1)\right)}\\\nonumber
&&\times \sinc\left( \pi\frac{\Delta}{c} \left(u_p l_q+v_p m_q+ w_p(\sqrt{1-l_q^2-m_q^2}-1)\right) \right).
\eeqn

Let $D$ be the number of pixels of the image in which the noise is calculated. Consider the construction of the image pixels given by $\widehat{\bf b}$ (size $D\times 1$)
\beq
\widehat{\bf b}=\left[ \begin{array}{c}
I(\tilde{l}_1,\tilde{m}_1)\\
I(\tilde{l}_2,\tilde{m}_2)\\
\ldots\\
I(\tilde{l}_D,\tilde{m}_D)
\end{array} \right]
\eeq
from the observed data $\tilde{\bf b}$. We can write this as
\beq \label{T}
\widehat{\bf b} =\widetilde{\bf T}_{f_0}^{\dagger} \widetilde{\bf b}
\eeq
where the elements in the matrix $\widetilde{\bf T}_{f_0}$ (size $M\times D$) are given as
\beq \label{Telem}
[\widetilde{\bf T}_{f_0}]_{pq}=e^{-j 2\pi \frac{f_0}{c}\left(u_p l_q+v_p m_q+ w_p(\sqrt{1-l_q^2-m_q^2}-1)\right)}
\eeq
and in the reconstruction, no smearing is assumed. Note also that the sets $\mathcal{L}$ and $\widetilde{\mathcal{L}}$ that denote the positions of the outlier sources and the positions of the pixels
\beq \label{pixelpos}
\mathcal{L}=\{(l_1,m_1),\ldots,(l_N,m_N)\},\ \ \widetilde{\mathcal{L}}=\{(\tilde{l}_1,\tilde{m}_1),\ldots,(\tilde{l}_D,\tilde{m}_D)\}
\eeq
have no relation to each other. In general the pixels coordinates $\widetilde{\mathcal{L}}$  where we calculate the differential noise, are on a regular grid.
Using (\ref{Tf}) and (\ref{T}), we get
\beq
\widehat{\bf b} =\widetilde{\bf T}_{f_0}^{\dagger} {\bf T}_{f_0} {\bf b}
\eeq
and the difference image at frequencies $f_1$ and $f_0$ can be given as
\beq
{\bf e}=(\widetilde{\bf T}_{f_1}^{\dagger} {\bf T}_{f_1}- \widetilde{\bf T}_{f_0}^{\dagger} {\bf T}_{f_0}) {\bf b}
\eeq
assuming the variation of ${\bf b}$ is negligible within this bandwidth.
The noise variance in the difference image is proportional to $\|{\bf e}\|^2$ and the average noise variance per pixel is $\|{\bf e}\|^2/D$ and the standard deviation is $\sqrt{|{\bf e}\|^2/D}$.

Considering a random distribution of outlier sources, we can also find the expected value of $\|{\bf e}\|^2$ as 
\beq \label{eeb}
E\{ \|{\bf e}\|^2 \}=E\{ {\bf e}^T {\bf e} \} = E\{ {\bf b}^T {\bf T}_{f_1 f_0} {\bf b} \}
\eeq
where ${\bf T}_{f_1 f_0}$ (size $N\times N$) is given by
\beq
{\bf T}_{f_1 f_0}={\mathrm{Re}}\left((\widetilde{\bf T}_{f_1}^{\dagger} {\bf T}_{f_1}- \widetilde{\bf T}_{f_0}^{\dagger} {\bf T}_{f_0})^H (\widetilde{\bf T}_{f_1}^{\dagger} {\bf T}_{f_1}- \widetilde{\bf T}_{f_0}^{\dagger} {\bf T}_{f_0}) \right)
\eeq
and note that ${\bf T}_{f_1 f_0}$ is symmetric positive semi-definite because of the factorization as above. We also consider ${\bf T}_{f_1 f_0}$ to be real because the sky image is real.

We find an upper and a lower bound for $E\{ \|{\bf e}\|^2 \}$ as 
\beq \label{vbound}
\frac{1}{N^2}E\{ \mid {\bf T}_{f_1 f_0} \mid \} E\{ |{\bf b}|^2 \} \le E\{ \|{\bf e}\|^2 \} \le  E\{ \mathrm{max}(\mathrm{diag}({\bf T}_{f_1 f_0})) \} E\{ |{\bf b}|^2 \}
\eeq
where $\mid {\bf T}_{f_1 f_0} \mid$ is the sum of all elements in ${\bf T}_{f_1 f_0}$ and $|{\bf b}|$ is the sum of all elements in ${\bf b}$. The diagonal entry with the highest magnitude in ${\bf T}_{f_1 f_0}$ is given by $\mathrm{max}(\mathrm{diag}({\bf T}_{f_1 f_0}))$. The proof and the underlying assumptions can be found in appendix \ref{proof1}.

We draw a few conclusions from (\ref{vbound}): (i) By properly selecting the $uv$ coverage and the frequencies $f_1$ and $f_0$ and also imaging weights (in the derivation natural weights are assumed) and image pixel sizes, we can change ${\bf  T}_{f_1 f_0}$. (ii) However, $| {\bf b} |^2$ is entirely determined by the intensities of the outlier sources and the only way to minimize $| {\bf b} |^2$ is by suppressing the beam sidelobe level or by subtracting the outlier sources, as we have done for CasA. (iii) We get the lowest value for the noise when all outlier sources have intensities that are equally distributed while we get the highest noise when there is one very bright source (see appendix \ref{proof1}). 

\begin{figure}[htbp]
\begin{center}
\epsfxsize=3.4in \leavevmode\epsfbox{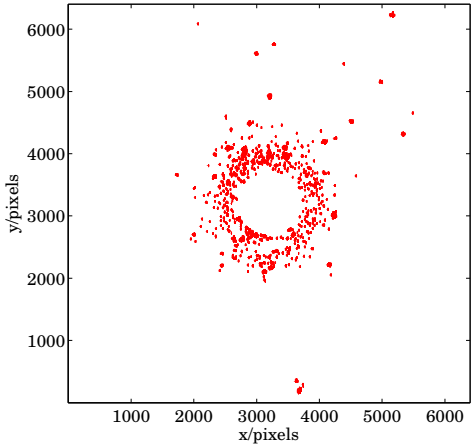}
\caption{Pixel locations outside the 10 degree FOV of the NCP image where outlier sources (flux $> 1.2$mJy) are detected. The image noise is about $0.3$mJy and we have selected $24000$ pixels out of $6400\times6400$ pixels.\label{ncp_outlier_grid}}
\end{center}
\end{figure}

In order to relate (\ref{vbound}) to NCP observations, we need to estimate $| {\bf b} |^2$ due to the outlier sources in the images. For the particular case of the NCP, we select the short baselines ($< 1200$ wavelengths) from the $uv$ coverage in Fig. \ref{uvcov}. The corresponding image of the FOV is shown in Fig. \ref{ncp_calib_sage_core}. With the same pixel size of 35$^{\prime\prime}$, we have made an image of $6400\times 6400$ pixels. From this image, we have selected all pixels that have a flux $> 1.2$mJy that are outside the FOV and we show the selected pixels in Fig. \ref{ncp_outlier_grid}.

\begin{figure}[htbp]
\begin{center}
\epsfxsize=3.4in \leavevmode\epsfbox{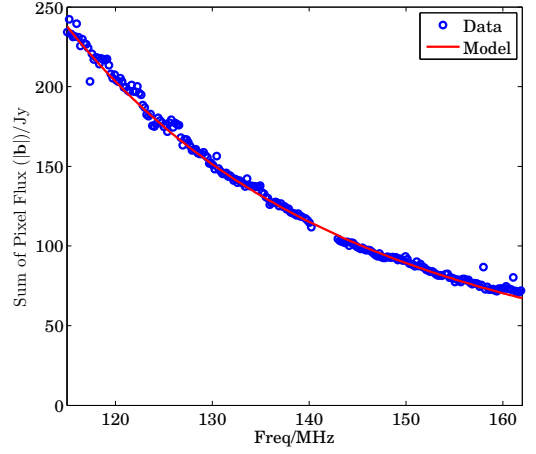}
\caption{The variation of $| {\bf b} |$ with frequency (circles). The model fitted to $| {\bf b} |$ is shown by the solid red line. The length of ${\bf b}$ is $24000$ and the pixels are chosen as shown in Fig. \ref{ncp_outlier_grid}.\label{ncp_outlier_noise}}
\end{center}
\end{figure}

We consider the vector ${\bf b}$ to be formed by the pixels selected in Fig. \ref{ncp_outlier_grid}. After forming this vector, we have calculated $| {\bf b} |$ for different frequencies (using images made at those frequencies) as shown in Fig. \ref{ncp_outlier_noise}. We have also fitted a model for $| {\bf b} |$ (the red line) which is
\beq \label{bnorm}
| {\bf b} |=89.32(\frac{f}{150\times 10^6})^{-3.687}+70.
\eeq

\begin{figure}[htbp]
\begin{center}
\epsfxsize=3.4in \leavevmode\epsfbox{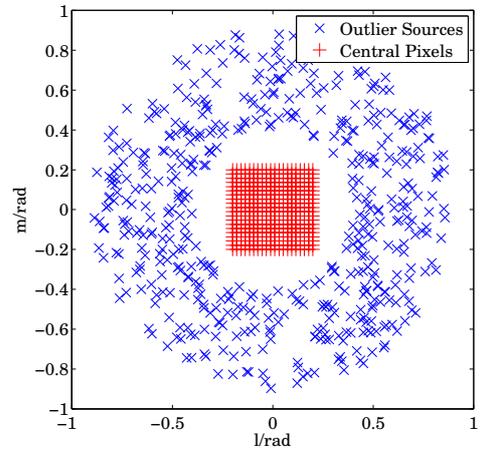}
\caption{Outlier source positions and the pixel grid positions on which the differential noise is calculated.\label{outlier_center}}
\end{center}
\end{figure}

Using the model for  $| {\bf b} |$ given by (\ref{bnorm}), we can approximate $E\{| {\bf b} |^2 \}$. Next, using this approximation, we simulate (\ref{vbound}), for various image differencing frequencies.  We simulate a sky consisting of $N=500$ outlier sources. Note that some of these sources will have intensities below the receiver noise level at any given frequency. But we make an important distinction that what matters is the cumulative effect of all the sources taken together, and not their individual contributions. The pixel grid for the simulation at the center has dimensions $100\times 100$, so $D=10000$.  The $uv$ coverage is chosen to be the core ($< 1200$ wavelengths) $uv$ coverage from Fig. \ref{uvcov}. For a longest baseline of 1200 wavelengths, the resolution is about 3$^\prime$ and the pixel size is chosen to be large enough (10$^\prime$) so that we are not dominated by sampling errors \citep{SAM2010} and that the noise is not correlated between pixels. The geometry of one realization of the simulation is shown in Fig. \ref{outlier_center}. Note that the positions of the outlier sources are randomly varied in each realization.

We have done simulations for three different values of smearing bandwidth, $\Delta=$200~kHz, $\Delta=$20~kHz, and $\Delta=$2~kHz. The bandwidth where the difference is taken is also equal to $\Delta$. For each value of $\Delta$, we generated $100$ different sky realizations as in Fig.  \ref{outlier_center}.

\begin{figure}[htbp]
\begin{minipage}{1.00\linewidth}
\begin{center}
\epsfxsize=3.0in \leavevmode\epsfbox{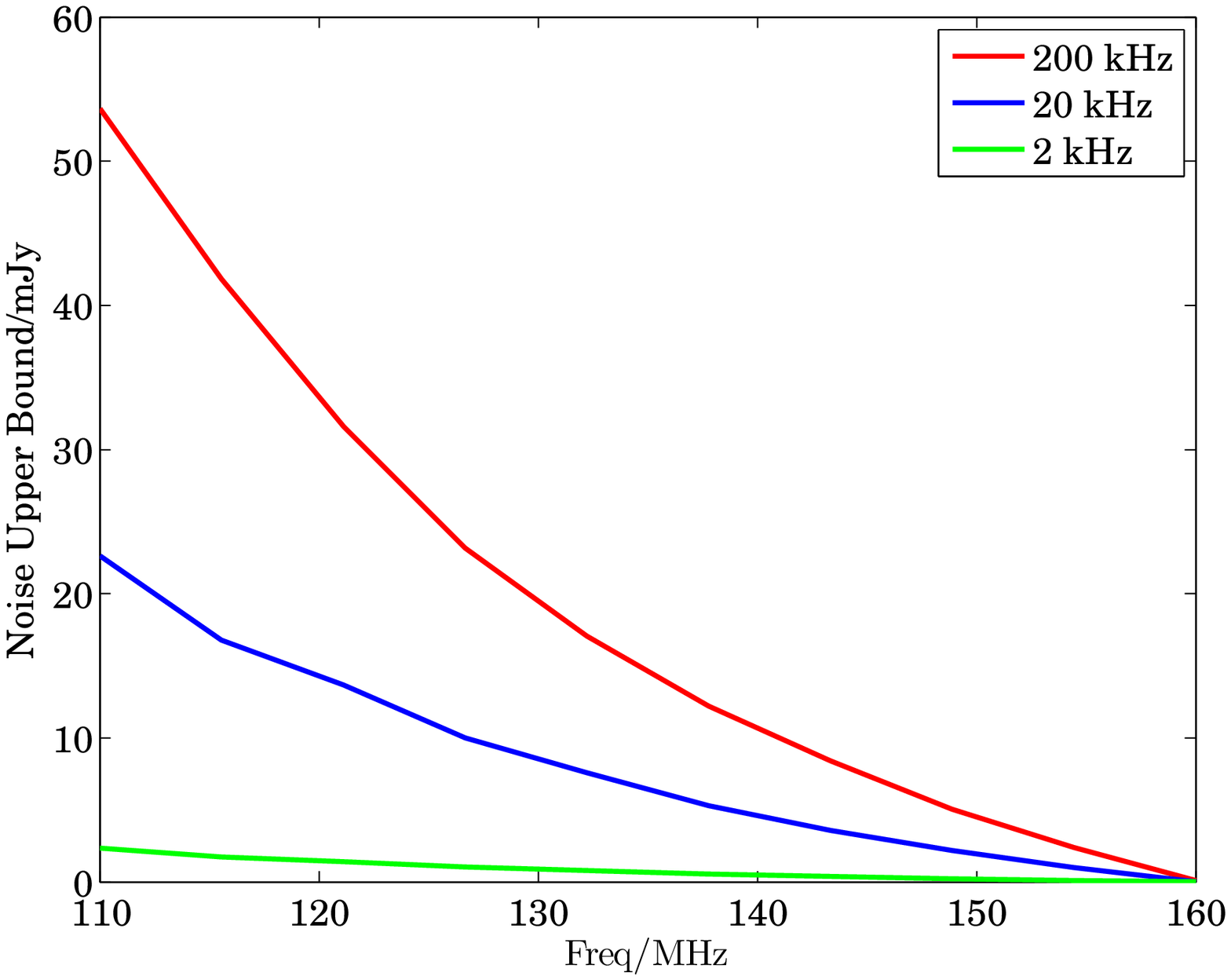}
\vspace{0.1cm}\centerline{(a)}\smallskip
\end{center}
\begin{center}
\epsfxsize=3.0in \leavevmode\epsfbox{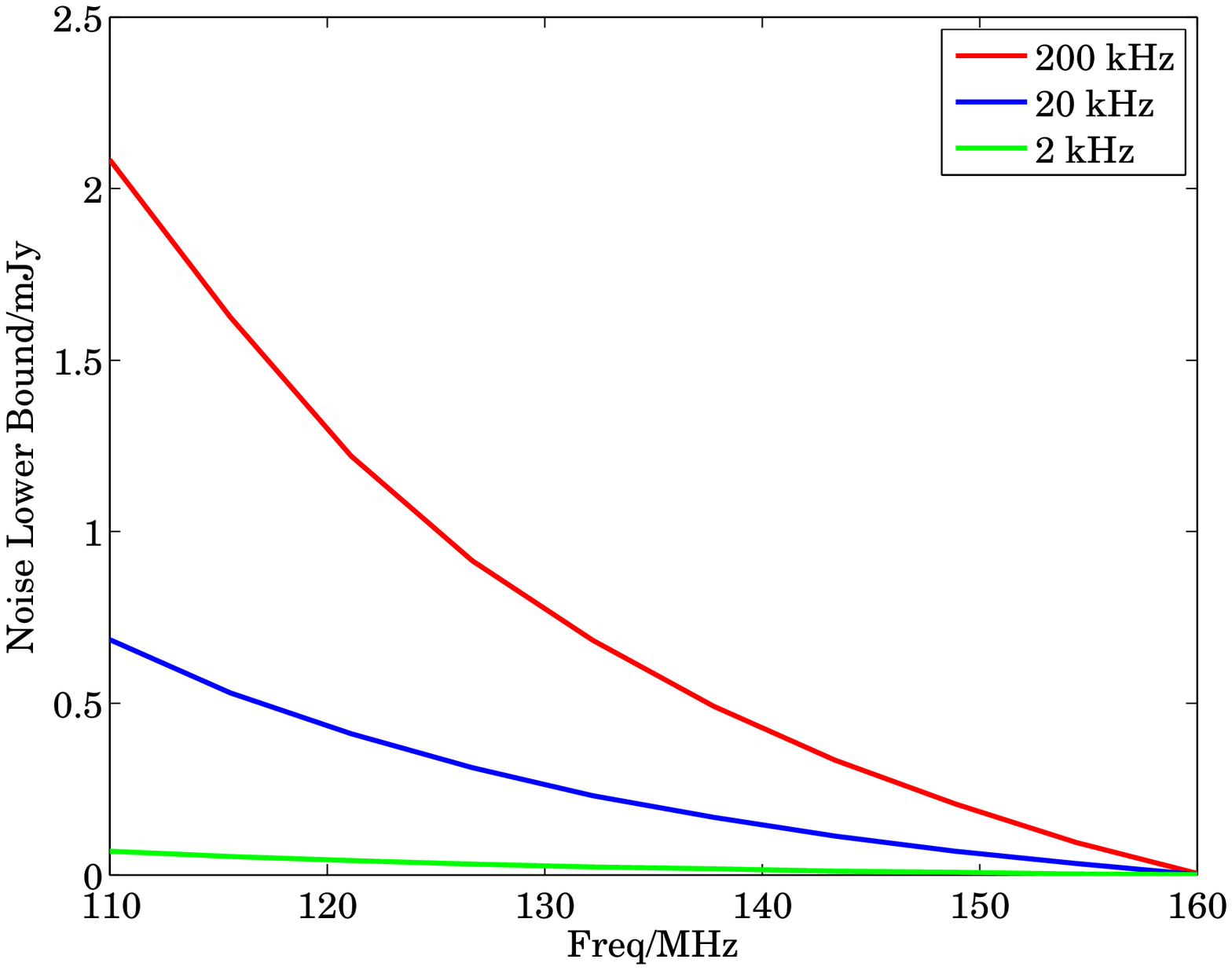}
\vspace{0.1cm}\centerline{(b)}\smallskip
\end{center}
\end{minipage}
\caption{The noise standard deviation due to outlier source for different values of smearing (and differencing) bandwidth. The upper bound is given in (a) and the lower bound is given in (b). As the smearing bandwidth decreases, the bounds on the noise are also reduced.\label{outlier_noise}}
\end{figure}

We have shown the upper and lower bounds for the standard deviation of the noise due to outlier sources per pixel in Fig. \ref{outlier_noise}. We assume the noise contribution due to outlier sources at the highest frequency (160 MHz) is zero. As expected, as the value of $\Delta$ decreases, the noise due to far away sources is also reduced. So for low values of $\Delta$, the result will be dominated by the receiver noise. In future processing, we intend to use $\Delta=$20~kHz and lower.

\section{Discussion/Conclusions\label{sec:conc}}
We have presented the first deep imaging results made with LOFAR that reach a noise level of about 100 $\mu$Jy/PSF both in total intensity and in polarization. Note that for calculation of the noise in polarization, we have only used one night of data while for calculation of the noise in total intensity, we have used all three nights of data listed in Table \ref{tab:results}. Using all three nights of data for polarization analysis requires correction for the effects due to differential Faraday rotation which varies from night to night and will be done in future processing. These results act as a precursor of results with more dedicated EoR observations that are about to begin with LOFAR. 

We briefly discuss the differences between previous WSRT observations of the NCP (72 hours, 14 MHz)  and the LOFAR observations (10 hours effective, 48 MHz) presented in this paper. Due to having only a 2.7 km longest baseline, the WSRT observations were limited by classical source confusion in the continuum.  Moreover, the elevated location of the WSRT receivers, and the interference from the building,  made them much more prone to RFI. The WSRT observations also used only 14 stations while LOFAR had about 40 stations  Furthermore, the data processing capabilities are significantly better than what was available to process WSRT observations. Therefore, despite the much longer integration time used in the WSRT observations we have obtained images that are much better in terms of resolution as well as noise. In fact, we have presented the deepest interferometric images ever made at this frequency range, to this date. The observed noise levels -- in high resolution images, linear polarization and at the edges of low resolution wide field images,  are within a factor of 1.4 from the theoretical noise that can be achieved based on the nominal system equivalent flux density and the effective integration time.

Nonetheless, we can see limitations in the current results that  can be attributed to the choices made in our processing: 
\begin{itemize}
\item We have only a partially correct sky model that contains about 500 discrete sources. However, we see much more than this in our images but at present we are limited by the resolution that is only about 12$^{\prime\prime}$. We therefore intend to use more data with longer baselines (even international LOFAR stations) to update our sky model. 
\item We have thus far not seen (and ignored) the effect of (spatial) differential Faraday rotation (DFR) due to the ionosphere in our data. Note that even though we calibrate for full Jones matrices using {\tt SAGECal}, we do not correct the data using the solutions. However, we shall soon incorporate taking DFR into account in our calibration \citep{DFR}. We believe this will be crucial when we get data with significantly longer (80~km) baselines. 
\item In the processing of our data we were limited by computational resources. We have therefore invested significant effort in optimizing the performance of our calibration pipeline using  GPUs to accelerate our processing. This will enable us to employ sophisticated algorithms yielding better results. For instance, with a slightly enhanced version of {\tt SAGECal} \citep{SAGECAL}, we intend to process data at a finer spectral resolution, therefore eliminating the bandwidth smearing visible at the edge of the FOV in our images. This will also decrease the effect of outlier sources, as discussed in section \ref{sec:outlier}. 
\end{itemize}
In summary, we do not see any major obstacles to prevent us from going considerably deeper through much longer multiple-nights integration. In forthcoming publications we will investigate the properties of the sources detected in the images,  make a detailed study of effects due to the Galactic foreground and determine limits on signals from the Epoch of Reionization.  

\section{Acknowledgements}
LOFAR, designed and constructed by ASTRON, has facilities in several countries,
that are owned by various parties (each with their own funding sources), and
that are collectively operated by the International LOFAR Telescope (ILT)
foundation under a joint scientific policy.
Chiara Ferrari acknowledges financial support by the {\it “Agence
Nationale de la Recherche”} through grant ANR-09-JCJC-0001-01.

\bibliographystyle{aa}
\bibliography{ncpref}
\begin{appendix}
\section{Proof of (\ref{vbound})\label{proof1}}
\subsection{Mathematical background}
Consider a vector of real numbers ${\bf x}$ of size $N\times 1$ with $x_i$ denoting the $i$-th element,
\beq
{\bf x}=[x_1,x_2,\ldots,x_N]^T.
\eeq
We can order the elements in ${\bf x}$ in descending order and construct a new vector
\beq
[{\bf x}]=[x_{[1]},x_{[2]},\ldots,x_{[N]}]^T
\eeq
where $x_{[1]}\ge x_{[2]} \ge \ldots \ge x_{[N]}$. Given two vectors of real numbers, ${\bf x}$ and ${\bf y}$ of size $N\times 1$, we say ${\bf x}$ is majorized by ${\bf y}$, or ${\bf x} \prec {\bf y}$ when 
\beqn \label{majorprop}
&&\sum_{i=1}^k x_{[i]} \le \sum_{i=1}^k y_{[i]},\ 1\le k \le N,\ \mathrm{and}\\\nonumber
&&\sum_{i=1}^N x_{[i]} = \sum_{i=1}^N y_{[i]}
\eeqn
are satisfied \citep{MOA}.

Consider the quadratic form
\beq
\phi({\bf x}) = {\bf x}^T {\bf A} {\bf x}
\eeq
where ${\bf A}$ is an $N\times N$ real symmetric matrix. If the elements in ${\bf A}$ satisfy
\beq \label{matprop}
\sum_{j=1}^{i} ([{\bf A}]_{k,j} -[{\bf A}]_{k+1,j}) \ge 0,\ i=1,\ldots,N,\ k=1,\ldots,N-1
\eeq
then we say $\phi({\bf x})$ is Schur-convex \citep{MOA}. 

If $\phi({\bf x})$ is Schur-convex and with two vectors of positive real numbers, ${\bf x}$ and ${\bf y}$, with  ${\bf x} \prec {\bf y}$, then
\beq \label{phi}
\phi({\bf x}) \le \phi({\bf y}).
\eeq
\subsection{Assumptions}
Using conditional mean, we rewrite (\ref{eeb}) as
\beq \label{cmean}
E\{ {\bf b}^T {\bf T}_{f_1 f_0} {\bf b} \} = E_{| {\bf b} |}\left\{ E\{\ \  \frac{{\bf b}^T}{|{\bf b}|} {\bf T}_{f_1 f_0} \frac{{\bf b}}{|{\bf b}|} |{\bf b}|^2 \ \  \mid \ \  | {\bf b} | \ \ \} \right\}.
\eeq
where $E_{| {\bf b} |}\{.\}$ is expectation with respect to $| {\bf b} |$, i.e. the sum of all elements in ${\bf b}$.

We make to following assumptions in order to proceed:
\begin{itemize}
\item The source fluxes (elements in ${\bf b}$) and their positions are both random variables. We assume that the probability distribution for the fluxes is statistically independent of the probability distribution for the source positions. In other words, the statistical properties of ${\bf b}$ in (\ref{cmean}) are independent of the statistical properties of ${\bf T}_{f_1 f_0}$.
\item We assume ${\bf T}_{f_1 f_0}$ to satisfy the properties given in (\ref{matprop}). More precisely, we assume that there exists a permutation of the rows and the columns of ${\bf T}_{f_1 f_0}$ to satisfy this condition. Since we did not impose any ordering on the pixel positions in (\ref{pixelpos}) this permutation is feasible. We do not need to find this permutation since the end result does not depend on it. But we assume that ${\bf T}_{f_1 f_0}$ is permuted such that the largest diagonal entry is at the $1$-st row and the $1$-st column.
\end{itemize}
\subsection{Derivation of bounds}
Let ${\bf z}=\frac{{\bf b}}{|{\bf b}|}$ and we see that ${\bf z}$ is a vector of $N$ positive elements that add up to $1$, i.e. $|{\bf z}|=1$. Consider the following vectors
\beqn
\underline{\bf z}&=&[1/N,1/N,\ldots,1/N]^T,\ \ \mathrm{and} \\\nonumber
\overline{\bf z}&=&[1,0,0,\ldots,0]^T.
\eeqn
We see that from the properties of (\ref{majorprop}), 
\beq \label{zed}
\underline{\bf z} \prec {\bf z} \prec \overline{\bf z}.
\eeq
Together with (\ref{phi}) and (\ref{zed}), and the assumptions on ${\bf T}_{f_1 f_0}$, we get
\beq \label{ineq1}
\underline{\bf z}^T {\bf T}_{f_1 f_0} \underline{\bf z} \le {\bf z}^T {\bf T}_{f_1 f_0} {\bf z} \le \overline{\bf z}^T {\bf T}_{f_1 f_0} \overline{\bf z}.
\eeq
We can simplify (\ref{ineq1}) as
\beq \label{ineq2}
\frac{1}{N^2}|{\bf T}_{f_1 f_0}| \le {\bf z}^T {\bf T}_{f_1 f_0} {\bf z} \le \mathrm{max}(\mathrm{diag}({\bf T}_{f_1 f_0}))
\eeq
where $|{\bf T}_{f_1 f_0}|$ is the sum of all elements in ${\bf T}_{f_1 f_0}$ and the diagonal entry with the highest magnitude of ${\bf T}_{f_1 f_0}$ (at row $1$ and column $1$) is given by $\mathrm{max}(\mathrm{diag}({\bf T}_{f_1 f_0}))$.

Substituting (\ref{ineq2}) back to (\ref{cmean})  and considering the independence of ${\bf b}$ and ${\bf T}_{f_1 f_0}$, we get (\ref{vbound}).

One additional point to note is that the lower bound is obtained when ${\bf z}=\underline{\bf z}$, i.e. when the elements in ${\bf b}$ are uniformly distributed with equal values. Moreover, the upper bound is obtained when ${\bf z}=\overline{\bf z}$, i.e. the total value of the elements in ${\bf b}$ is concentrated at one element. Reverting back to the fluxes of the sources, the lower bound is obtained when all sources have equal fluxes while the upper bound is obtained when only one source has the total flux (or when there is one very bright source).

\end{appendix}

\end{document}